\documentclass[12pt]{article}
\usepackage{jheppub}

\pdfoutput=1

\usepackage[table]{xcolor}
\usepackage{amsmath,bbm,array,amsfonts,graphicx,wrapfig,lscape,float,mathtools,multirow,longtable}
\usepackage{stackrel}
\usepackage[all]{xy}
\usepackage{subcaption}

\newcommand{\be}{\begin{equation}}
\newcommand{\ee}{\end{equation}}
\newcommand{\beq}{\begin{equation}}
\newcommand{\beql}[1]{\begin{equation}\label{#1}}
\newcommand{\eeq}{\end{equation}}
\newcommand{\ba}{\begin{array}}
\newcommand{\ea}{\end{array}}
\newcommand{\bea}{\begin{eqnarray}}
\newcommand{\beal}[1]{\begin{eqnarray}\label{#1}}
\newcommand{\eea}{\end{eqnarray}}
\newcommand{\ben}{\begin{enumerate}}
\newcommand{\een}{\end{enumerate}}
\newcommand{\bean}{\begin{eqnarray*}}
\newcommand{\eean}{\end{eqnarray*}}
\newcommand{\eref}[1]{(\ref{#1})}
\newcommand{\sref}[1]{\S\ref{#1}}
\newcommand{\tref}[1]{Table~\ref{#1}}
\newcommand{\nn}{\nonumber}

\newcommand{\fref}[1]{Figure \ref{#1}}
\newcommand{\btab}[1]{\begin{tabular}{#1}}
\newcommand{\etab}{\end{tabular}}

\newcommand{\comment}[1]{}

\newcommand{\qed}{\nobreak \ifvmode \relax \else
      \ifdim\lastskip<1.5em \hskip-\lastskip
      \hskip1.5em plus0em minus0.5em \fi \nobreak
      \vrule height0.75em width0.5em depth0.25em\fi}

\definecolor{darkspringgreen}{rgb}{0.09, 0.45, 0.27}
\definecolor{forestgreen}{rgb}{0.13, 0.55, 0.13}

\usepackage{array}

\newcolumntype{C}[1]{>{\centering\let\newline\\\arraybackslash\hspace{0pt}}m{#1}}

\title{4d Crystal Melting, Toric Calabi-Yau 4-Folds and Brane Brick Models} 

\author[a,b,c]{Sebasti\'an Franco}

\affiliation[a]{Physics Department, The City College of the CUNY\\
	160 Convent Avenue, New York, NY 10031, USA}
\affiliation[b]{Physics Program and \textsuperscript{$c$}Initiative for the Theoretical Sciences\\
	The Graduate School and University Center, The City University of New York\\
	365 Fifth Avenue, New York NY 10016, USA}

\emailAdd{sfranco@ccny.cuny.edu}

\abstract{We introduce a class of 4-dimensional crystal melting models that count the BPS bound state of branes on toric Calabi-Yau 4-folds. The crystalline structure is determined by the brane brick model associated to the Calabi-Yau 4-fold under consideration or, equivalently, its dual periodic quiver. The crystals provide a discretized version of the underlying toric geometries. We introduce various techniques to visualize crystals and their melting configurations, including 3-dimensional slicing and Hasse diagrams. We illustrate the construction with the D0-D8 system on $\mathbb{C}^4$. Finally, we outline how our proposal generalizes to arbitrary toric CY 4-folds and general brane configurations.}

\begin{document}

\maketitle

\section{Introduction}

Probing Calabi-Yau (CY) singularities with D-branes is a fruitful approach for engineering quantum field theories in various dimensions (see e.g. \cite{Klebanov:1998hh,Morrison:1998cs,Klebanov:2000hb,Aldazabal:2000sa,Verlinde:2005jr}). For toric CY's, the gauge theories are further endowed with beautiful combinatorial structures. A paradigmatic example is given by the $4d$ $\mathcal{N}=1$ gauge theories on D3-branes probing toric CY 3-folds, which are captured by {\it brane tilings} \cite{Hanany:2005ve,Franco:2005rj,Franco:2005sm}. Brane tilings have significantly simplified the connection between the gauge theories and the corresponding CY 3-folds, becoming standard tools with applications that range from string phenomenology to integrable systems.

Building on the seminal work \cite{Okounkov:2003sp,Iqbal:2003ds}, it was realized that the BPS spectrum of D-branes on a toric CY 3-fold is captured by a statistical model of {\it crystal melting} \cite{Ooguri:2009ijd,Ooguri:2009ri} (see also \cite{Szendroi:2007nu,Mozgovoy:2008fd} for important earlier ideas). Remarkably, the crystalline structure underlying these models is given by the brane tiling, or equivalently the dual periodic quivers, corresponding to the CY 3-fold. 

In recent years, a program similar to the one that lead to the discovery brane tilings has focused on understanding the $2d$ $(0,2)$ gauge theories on D1-branes probing toric CY 4-folds \cite{Franco:2015tna}. This program culminated with the introduction of {\it brane brick models}, a new class of type IIA brane configurations that are connected to the D1-branes at the singular CY 4-folds by T-duality \cite{Franco:2015tya}. Very much like their brane tiling precursors, brane brick models have trivialized the correspondence between $2d$ gauge theories and toric CY 4-folds.
We refer the interested reader to \cite{Franco:2016nwv,Franco:2016qxh,Franco:2016fxm,Franco:2017cjj,Franco:2018qsc,Franco:2019bmx,Franco:2020avj,Franco:2022iap,Franco:2022gvl,Franco:2022isw,Franco:2023tyf} for several further developments. 

Lately, Nekrasov introduced the {\it Magnificent Four}, a statistical model whose random variables are solid partitions \cite{Nekrasov:2017cih,Nekrasov:2018xsb,Nekrasov:2023nai}. The model computes the refined index of a system of D0-branes in the presence of D8-$\overline{\rm{D8}}$ system in $\mathbb{C}^4$, with a B-field.\footnote{See also \cite{Kimura:2023bxy} for interesting recent developments.}

Motivated by these recent developments, in this paper we introduce crystal melting models that capture the BPS bound states of D-branes on toric CY 4-folds. For general toric CY 4-folds, these models might involve D0/D2/D4/D6/D8-brane charges. The crystal is 4-dimensional and its crystalline structure is determined by the brane brick model (or, equivalently, its dual periodic quiver) associated to the CY 4-fold under consideration. Beautifully, but perhaps not surprisingly, brane brick models, which are instrumental in connecting quivers on D-branes to toric CY 4-folds, play a central role in the crystal melting models.

This paper is organized as follows. Section \sref{section_CY4_quivers_BBMs} presents a brief discussion of the $2d$ gauge theories on D1-branes probing toric CY 4-folds and their description in terms of brane brick models. Section \sref{section_C4} discusses the case of $\mathbb{C}^4$, which we will use thorough the paper to illustrate our ideas. Section \sref{section_combinatorics_BBMs} reviews important combinatorial objects associated to brane brick models, such as brick matchings and the oriented surfaces that result from subtracting them. Section \sref{section_height_function} discusses the concept of height function in brane brick models. Section \sref{section_model_crystal_melting_CY4_quiver} introduces a statistical model of crystal melting for toric CY 4-folds, focusing on $\mathbb{C}^4$. Section \sref{section_crystal_C4} constructs the crystal for $\mathbb{C}^4$ and initiates its investigation. It introduces fixed depth slicing as a useful approach for visualizing $4d$ crystals. Section \sref{section_exploring_the_crystal} continues with the exploration of the $\mathbb{C}^4$ crystal, introducing Hasse diagrams, a powerful tool to study crystals and melting configurations. In preparation for an implementation of the crystal model in terms of brane brick models, Section \sref{section_appetizer} presents related ideas for CY 3-folds and brane tilings. Section \sref{section_crystal_melting_BBM} reformulates the crystal melting model for $\mathbb{C}^4$ in terms of the corresponding brane brick model. Section \sref{section_crystal_general_CY4} outlines how our ideas extend to arbitrary toric CY 4-folds and general brane configurations. In Section \sref{section_conclusions}, we present our conclusions and outline various directions for future research.

\bigskip

\noindent {\bf Note:} While this paper was ready for submission, \cite{Galakhov:2023vic} appeared. Besides the common subject, there seems to be minor overlap with our work.

\section{Toric CY$_4$'s, $2d$ $(0,2)$ quivers and brane brick models}

\label{section_CY4_quivers_BBMs}

\begin{figure}[h]
	\centering
	\includegraphics[height=3cm]{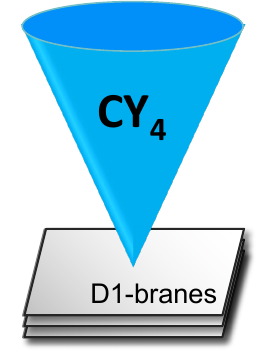}
\caption{D1-branes probing a CY$_4$.}
	\label{D1s_over_CY4}
\end{figure}

Consider a Type IIB setup of D1-branes probing a toric CY$_4$ singularity, as schematically shown in \fref{D1s_over_CY4}. The effective low energy theory on the worldvolume of the D1-branes is a $2d$ $(0,2)$ gauge theory.\footnote{D1-branes on generic CY$_4$'s preserve $(0,2)$ SUSY. Non-chiral SUSY enhancement occurs when the putative CY$_4$ contains $\mathbb{C}$ factors; $\mathbb{C}^4$, CY$_2\times\mathbb{C}^2$, CY$_3\times \mathbb{C}$ preserve $(8,8)$, $(4,4)$, $(2,2)$ SUSY, respectively. Chiral enhancement to $(0,4)$ SUSY arises from $\mathrm{CY}_2\times \mathrm{CY}_2$. Further chiral enhancement to $(0,6)$ or $(0,8)$ is possible for particular orbifold geometries.} 

Brane brick models are obtained from D1-branes at toric CY$_4$ singularities by T-duality. A brane brick model is a Type IIA brane configuration consisting of D4-branes wrapping a 3-torus $\mathbb{T}^3$ and suspended from an NS5-brane that wraps a holomorphic surface $\Sigma$ intersecting with $\mathbb{T}^3$ as summarized in \tref{Brane brick-config}. The holomorphic surface $\Sigma$ is the zero locus of the Newton polynomial of the toric $\text{CY}_4$. The $2d$ gauge theory lives on the two directions $(01)$ common to all the branes. The $(246)$ directions are compactified on a $\mathbb{T}^3$. Most of the important gauge theory information is captured by a tropical limit, i.e. a skeleton, of this configuration. For this reason, such skeleton is also often referred to as the brane brick model. 

\begin{table}[ht!!]
\centering
\begin{tabular}{l|cccccccccc}
\; & 0 & 1 & 2 & 3 & 4 & 5 & 6 & 7 & 8 & 9 \\
\hline
$\text{D4}$ & $\times$ & $\times$ & $\times$ & $\cdot$ & $\times$ & $\cdot$ & $\times$ & $\cdot$ & $\cdot$ & $\cdot$  \\
$\text{NS5}$ & $\times$ & $\times$ & \multicolumn{6}{c}{----------- \ $\Sigma$ \ ------------} & $\cdot$ & $\cdot$ \\
\end{tabular}
\caption{Brane brick model configuration.}
\label{Brane brick-config}
\end{table}

Brane brick models are dual to periodic quivers on $\mathbb{T}^3$. Both objects encode all the necessary information for writing the Lagrangian of the $2d$ $(0,2)$ quiver gauge theories on the worldvolume of D1-branes probing toric CY 4-folds. Namely, they summarize not only the quivers, but also their $J$- and $E$-terms. The dictionary connecting brane brick models to $2d$ $(0,2)$ gauge theories is summarized in Table \ref{tbrick}.

\begin{table}[H]
\centering
\resizebox{\hsize}{!}{
\begin{tabular}{|l|l|l|}
\hline
{\bf Brane Brick Model} \ \ &  {\bf Gauge Theory} \ \ \ \ \ \ \  & {\bf Periodic Quiver} \ \ \ 
\\
\hline\hline
Brick  & Gauge group & Node \\
\hline
Oriented face  & Bifundamental chiral field & Oriented (black) arrow 
\\
between bricks $i$ and $j$ & from node $i$ to node $j$  & from node $i$ to node $j$ \\
\hline
Unoriented square face  & Bifundamental Fermi field & Unoriented (red) line \\
between bricks $i$ and $j$ & between nodes $i$ and $j$ & between nodes $i$ and $j$  \\
\hline
Edge  & $J$- or $E$-term coupling & Plaquette encoding \\ 
& & a $J$- or an $E$-term \\
\hline
\end{tabular}
}
\caption{
Dictionary between brane brick models and $2d$ gauge theories.
\label{tbrick}
}
\end{table}

Various consistency conditions of the $2d$ $(0,2)$ gauge theory, such as anomaly cancellation and the trace condition are guaranteed by structural properties of brane brick models (see e.g. \cite{Franco:2015tya,Franco:2021elb}).

We refer the reader to \cite{Franco:2015tna,Franco:2015tya,Franco:2016nwv,Franco:2016qxh,Franco:2021elb} for further details. Brane brick models reduce the computation of the underlying CY$_4$ geometry starting from the gauge theory to a combinatorial problem, which is based on a generalization of perfect matchings to be discussed in Section \sref{section_combinatorics_BBMs}. Conversely, several efficient algorithms for determining the brane brick models for a given geometry have been developed \cite{Franco:2015tna,Franco:2016qxh,Franco:2016fxm,Franco:2018qsc,Franco:2020avj,Closset:2017yte,Closset:2018axq}.

\section{The simplest example: $\mathbb{C}^4$}

\label{section_C4}

\subsection{Quiver theory}

\begin{figure}[h]
	\centering
	\includegraphics[height=4cm]{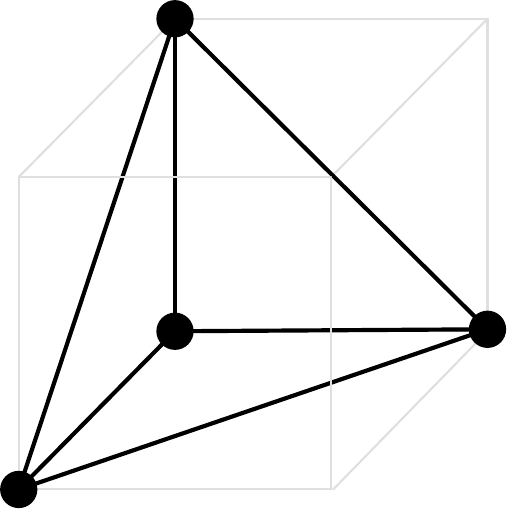}
\caption{Toric diagram of $\mathbb{C}^4$.}
	\label{toric_C4}
\end{figure}

Let us consider the gauge theory on D1-branes on $\mathbb{C}^4$, whose toric diagram is shown in \fref{toric_C4}. 

\begin{figure}[h]
	\centering
	\includegraphics[height=3.5cm]{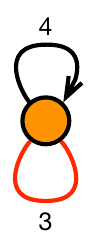}
\caption{Quiver diagram for $N$ D1-branes over $\mathbb{C}^4$. It consists of a single $U(N)$ gauge node, four adjoint chiral fields (shown in black) and three Fermi fields (shown in red).}
	\label{quiver_C4}
\end{figure}

This theory can be obtained by dimensional reduction of $4d$ $\mathcal{N}$=4 SYM. The resulting $2d$ theory has $(8,8)$ SUSY. In $(0,2)$ language, the theory contains a vector multiplet associated with a single $U(N)$ gauge group, four chiral fields ($X$, $Y$, $Z$ and $W$) and three Fermi fields ($\Lambda_i$, $i=1,2,3$), all transforming in the adjoint representation of the gauge group. This information is summarized in the quiver shown in \fref{quiver_C4}.

The corresponding $J$- and $E$-terms are as follows
\beq
\begin{array}{cccc}
& J & & E \\
\Lambda_1 : & Y \cdot Z - Z \cdot Y & \ \ \ \ & W \cdot X - X \cdot W  \\ 
\Lambda_2 : & Z \cdot X - X \cdot Z & \ \ \ \ & W \cdot Y - Y \cdot W  \\ 
\Lambda_3 : & X \cdot Y - Y \cdot X & \ \ \ \ & W \cdot Z - Z \cdot W  
\end{array}
\label{EJ_C^4}
\eeq

\subsection*{Adding flavor to the quiver with a D9-brane}

\begin{figure}[h]
	\centering
	\includegraphics[height=3.5cm]{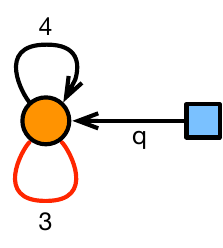}
\caption{The quiver diagram for $N$ D1-branes over $\mathbb{C}^4$. It consists of a single $U(N)$ gauge node, four adjoint chiral fields (shown in black) and three Fermi fields (shown in red).}
	\label{quiver_C4_flavor}
\end{figure}

We can add flavor fields to the quiver by introducing higher dimensional branes. Let us consider a single D9-brane spanning the two dimensions of the gauge theory plus the entire $\mathbb{C}^4$. The gauge theory is the same one as before, with the addition of a single chiral arrow $q$, as shown in \fref{quiver_C4_flavor}. The flavor node, shown in blue, represents the D9-brane. The field $q$ does not participate in any $J$- or $E$-terms.

\subsection{D0-branes on CY 4-folds}

In what follows, we will interpret the class of theories described by brane brick models as the supersymmetric quantum mechanics on D0-branes probing the corresponding toric CY 4-folds. The extra flavor introduced in the previous section corresponds to a D8-brane with $B$-field spanning $\mathbb{C}^4$. In practice, we simply remove the two field theory dimensions from the previous discussion (or from the typical brane brick model literature, which is typically about theories on D1-branes).

In this context, the chiral and Fermi fields of the unflavored theories are in the D0-D0 sector. The flavor introduced in the previous section corresponds to a D8-D0 field. The flavored quiver theory in the previous section is indeed the one underlying the Magnificent Four model. More general flavor branes will be discussed in Section \sref{section_crystal_general_CY4}.

\subsection{Brane brick model for $\mathbb{C}^4$}

\begin{figure}[h]
	\centering
	\includegraphics[height=5cm]{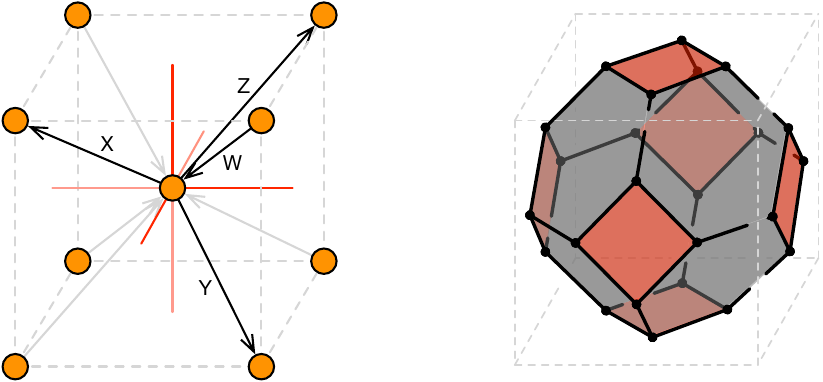}
\caption{Periodic quiver and dual brane brick model for $\mathbb{C}^4$.}
	\label{quiver_BBM_C4}
\end{figure}

\fref{quiver_BBM_C4} shows the periodic quiver and dual brane brick model for $\mathbb{C}^4$. Black and red faces in the brane brick model correspond to chiral and Fermi fields, respectively. To simplify the figure, the region shown in both cases is larger than a unit cell.

\bigskip

\begin{figure}[h]
	\centering
	\includegraphics[height=5.5cm]{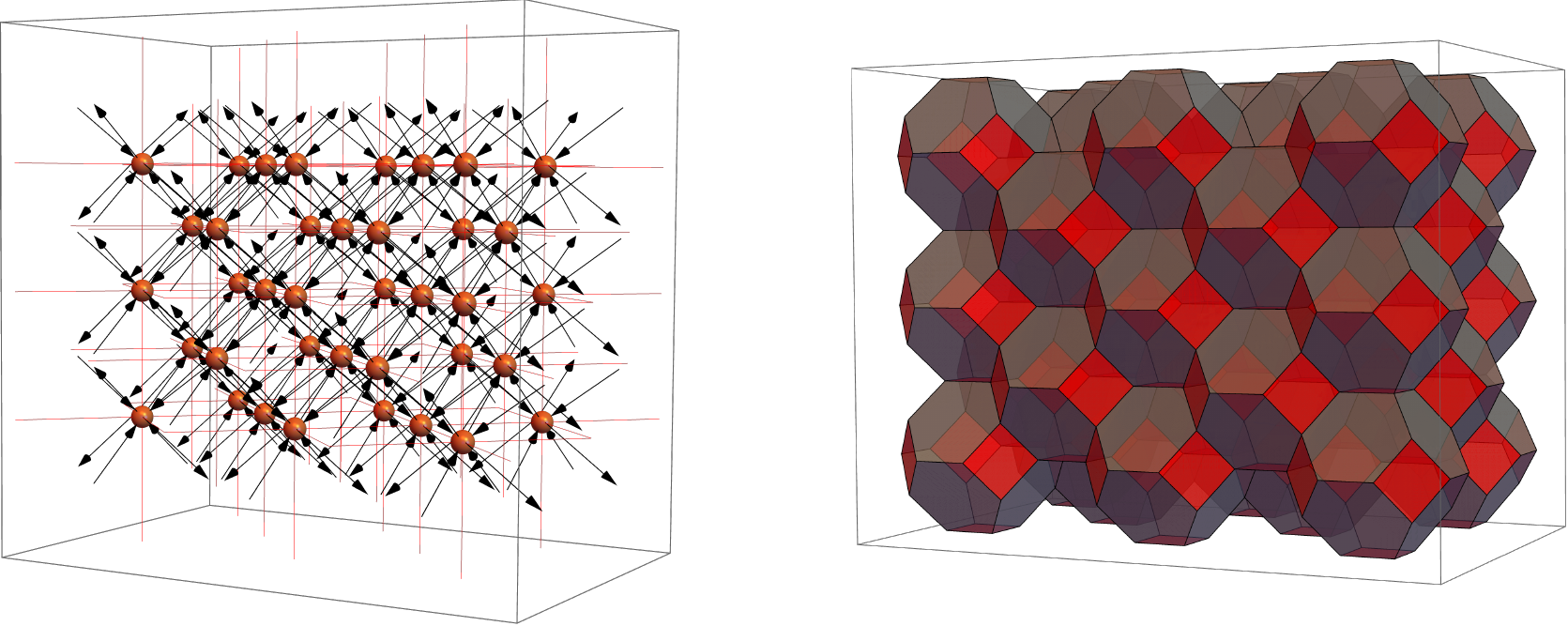}
\caption{Universal covers of the periodic quiver and brane brick model for $\mathbb{C}^4$.}
	\label{universal_cover_quiver_BBM}
\end{figure}

Throughout the paper, we will often consider the universal covers of the periodic quiver and the brane brick model. \fref{universal_cover_quiver_BBM} illustrates how the combination of several unit cells of each of them looks like.

\bigskip

\newpage

\section{The combinatorics of brane brick models}

\label{section_combinatorics_BBMs}

In this section we present a brief review of certain combinatorial objects associated to brane brick models. We refer the reader to \cite{Franco:2015tya} for further details.

\subsection{Brick matchings}

\label{section_brick_matchings}

{\it Brick matchings} are the brane brick model analogues of perfect matchings for dimer models. They play a central role in the connection between the quiver theories and the underlying CY 4-folds. Below we present three equivalent definitions for them.

\paragraph{Definition 1.}

Brick matchings can be defined through the $J$- and $E$-terms of the brane brick model which, for gauge theories on the worldvolume of D1-branes probing toric CY 4-folds, take the following binomial form 
\beal{es05a01}
\Lambda_{a} &~:~&
J_a = J_a^{+} - J_a^{-} 
\nn\\
\bar{\Lambda}_{a} &~:~&
E_a = E_a^{+} - E_a^{-}  
~,~
\eea
which is often referred to as the toric condition \cite{Franco:2015tna}. Here, the index $a$ runs over Fermi fields. To define the brick matchings, we complete the $J$- and $E$-terms into gauge invariants by multiplying them by the corresponding Fermi fields $\Lambda_{a}$ or conjugate Fermi fields $\bar{\Lambda}_{a}$.
This results in two pairs of monomial terms known as \textit{plaquettes} for every $(\Lambda_{a}, \bar{\Lambda}_{a})$-pair, 
\beal{es05a02}
\Lambda_{a} \cdot J_a^{+} ~,~
\Lambda_{a} \cdot J_a^{-} ~,~
\bar{\Lambda}_{a} \cdot E_a^{+} ~,~
\bar{\Lambda}_{a} \cdot E_a^{-} ~,~
\eea
where $J_a^{\pm}$ and $E_a^{\pm}$ indicate holomorphic monomial products of chiral fields. 
Given plaquettes, \textit{brick matchings} are defined as a special collection of chiral, Fermi and conjugate Fermi fields that cover every plaquette exactly once by satisfying the following conditions:
\begin{itemize}
\item The chiral fields in the brick matching cover the plaquettes $(\Lambda_{a} \cdot J_a^{+},~ \Lambda_{a} \cdot J_a^{-} )$ or the plaquettes $(\bar{\Lambda}_{a} \cdot E_a^{+},~ \bar{\Lambda}_{a} \cdot E_a^{-} )$ exactly once each. 

\item If the chiral fields in the brick matching cover the plaquettes $(\Lambda_{a} \cdot J_a^{+},~ \Lambda_{a} \cdot J_a^{-} )$, then $\bar{\Lambda}_{a}$ is included in the brick matching.

\item If the chiral fields in the brick matching cover the plaquettes $(\bar{\Lambda}_{a} \cdot E_a^{+},~ \bar{\Lambda}_{a} \cdot E_a^{-} )$, then $\Lambda_{a}$ is included in the brick matching. 
\end{itemize}

The chiral fields $X_m$ contained in brick matching $p_\mu$ can be summarized in a brick matching matrix $P$, whose entries take the following form,
\beal{es05a10}
P_{m\mu} =  \left\{ \ba{cc}
1 & ~~~X_m \in p_\mu \\
0 & ~~~X_m \notin p_\mu 
\ea
\right.
~.~
\eea
The Fermi field content of a brick matching can be reconstructed from knowledge of the chiral fields in it, so the latter is sufficient for determining it.\footnote{Interestingly, contrary to what happens for ordinary perfect matchings of brane tilings, brick matchings can have different numbers of chiral fields.} Moreover, only chiral fields are necessary for connecting with the underlying toric geometry. Including the (conjugate) Fermis is important when constructing surfaces by taking differences of perfect matchings, as it will be discussed in Section \sref{section_surfaces_from_brick_matchings}. If the Fermis were not present, the resulting surfaces would have holes at their locations.

Given the $P$-matrix, it is useful to express chiral fields in terms of brick matchings as follows
\beal{es05a11}
X_m = \prod_\mu p_\mu^{P_{m\mu}}
~.~
\eea
Remarkably, the combinatorial structure of brick matchings is such that the map \eref{es05a11} between chiral fields and perfect matching variables automatically satisfies the vanishing $J$- and $E$-terms conditions. This, in turns, leads to a one-to-one correspondence between brick matchings and GLSM fields in the toric description of the classical mesonic moduli space of the gauge theory. As such, perfect matching map to points in the toric diagram of the CY$_4$. There are various ways for determining the position in the toric diagram of a given perfect matching. They include: assigning them charges under the gauge symmetries and imposing vanishing $D$-terms, computing their intersections with the fundamental axes of the brane brick model unit cell, and using the slope of the height function \cite{Franco:2015tya}.

\bigskip

\paragraph{Definition 2.}

It is worth mentioning that there exist an alternative definition of brick matchings due to Richard Kenyon, which is identical the one for perfect matchings of brane tilings (see e.g. \cite{Franco:2005rj}):\footnote{We thank Richard Kenyon for private discussions leading to this insight. These conversations took place during a meeting of the NSF FRG in the Mathematical Sciences shared with the author, and benefitted from ideas from the other members of the group: Gregg Musiker, David Speyer and Lauren Williams.}
\begin{itemize}
\item A perfect matching $p$ is such that every vertex in the brane brick model 
is covered exactly once by a chiral face in $p$.
\end{itemize}

\bigskip

\paragraph{Definition 3: Brick matchings from chiral cycles.}

Let us consider the $J$- and $E$-terms associated to a Fermi field $\Lambda_{a}$. The product 
\begin{align}
J_{a}E_{a} = J^{+}_{a}E^{+}_{a} - J^{+}_{a}J^{-}_{a} - J^{-}_{a}E^{+}_{a} + J^{-}_{a}E^{-}_{a}
\label{chiral_cycles_BBM}
\end{align}
is a sum of four {\it chiral cycles}. From the point of view of the periodic quiver, chiral cycles are ``minimal" closed oriented loops of chiral fields.

We can alternatively define the chiral content of a brick matching as a collection of chiral fields that contains exactly one field from each of these chiral cycles for every Fermi field \cite{Franco:2019bmx}. It is easy to see that, according to this definition, a brick matching has two (not necessarily distinct) chiral fields from the $J$- and $E$-terms of a given Fermi field $\Lambda_{a}$, and either both of them belong to $J_{a}$ or both belong to $E_{a}$. Hence, it covers either both $J$-terms and we add $\bar{\Lambda}_{a}$ to it, or it covers only $E$-terms and we add $\Lambda_{a}$. With this completion with Fermi fields, this definition is clearly equivalent to the first one.

This definition of perfect matchings, combined with \eref{es05a11}, implies that all chiral cycles are equivalent on-shell, i.e. modulo $J$- and $E$-term relations. Every chiral cycle becomes equal to the product of all perfect matchings, when written in terms of these variables. Therefore, we can equate all these minimal closed cycles to a single variable that we will call $\omega$. While this might be rather expected in the case of $\mathbb{C}^4$, for which all chiral cycles are quartic products of chiral fields and the four chiral fields are symmetric, it is non-trivial for general toric CY 4-folds. This fact will become important in Section \sref{section_model_crystal_melting_CY4_quiver} when we construct a crystal from the quiver.

To conclude, let us mention that this definition of perfect matchings extends to a generalization of dimer models for toric CY$_{m+2}$ with arbitrary $m\geq 0$, which was introduced in \cite{Franco:2019bmx}. It is natural to expect that such {\it $m$-dimers} and their perfect matchings may be relevant to the extensions of crystal melting to higher dimensional CY's.

\subsection*{Brick matchings for $\mathbb{C}^4$}

\begin{figure}[ht!]
	\centering
	\includegraphics[height=2.5cm]{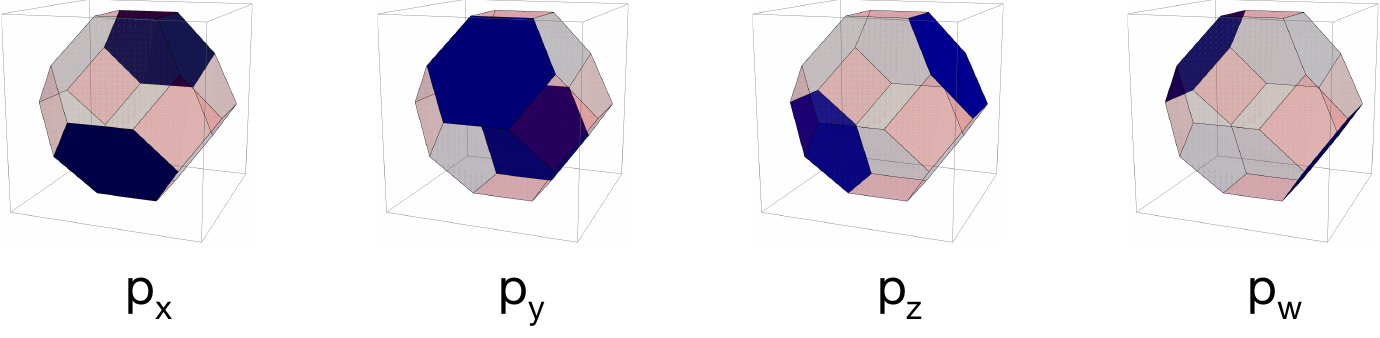}
\caption{The four brick matchings for $\mathbb{C}^4$.}
	\label{pms_C4}
\end{figure}

The theory for $\mathbb{C}^4$ has four brick matchings, which are in one-to-one correspondence with the chiral fields associated to the four complex directions. They are presented in \fref{pms_C4}. We only show the chiral fields in them since, as mentioned earlier, this information is sufficient for determining the Fermi content.

The $P$-matrix summarizing the perfect matchings is therefore
\beq
P=\left(\begin{array}{c|cccc} 
& \ \ p_x \ \ & \ \ p_y \ \ & \ \ p_z \ \ & \ \ p_w \ \ \\ \hline 
X & 1 & 0 & 0 & 0 \\
Y & 0 & 1 & 0 & 0 \\
Z & 0 & 0 & 1 & 0 \\
W & 0 & 0 & 0 & 1
\end{array}\right) \, .
\eeq

\begin{figure}[ht!]
	\centering
	\includegraphics[height=8cm]{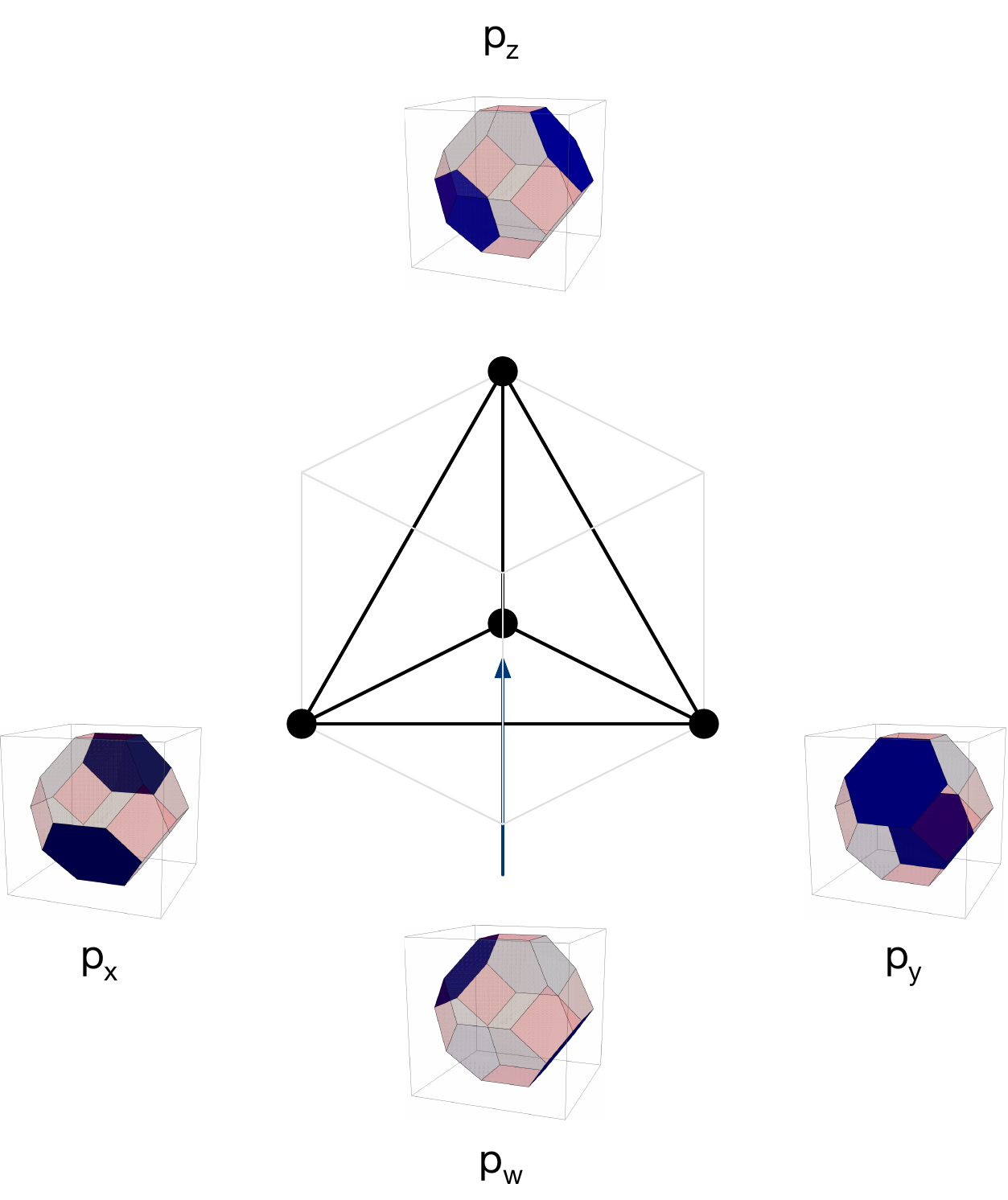}
\caption{Correspondence between brick matchings for $\mathbb{C}^4$ and points in its toric diagram.}
	\label{toric_C4_with_pms}
\end{figure}

\fref{toric_C4_with_pms} shows the correspondence between these brick matchings and points in the toric diagram of $\mathbb{C}^4$.

\subsection{Oriented surfaces from brick matchings}

\label{section_surfaces_from_brick_matchings}

The difference between two brick matching $p_1-p_2$ results in the (disjoint union of) oriented surfaces on the brane brick model or its universal cover. This difference is defined as follows:

\begin{itemize}
\item The orientation of every face in a perfect matching is given by the orientation of the corresponding bifundamental (or adjoint) fields.
\item The orientations of faces in $p_2$ are reversed.
\item Faces contained in both perfect matchings are combined with opposite orientations and hence disappear from the final result.
\item The orientation of the resulting surface is determined by the orientation of the chiral fields it contains.
\end{itemize}

Below we present two typical examples of surfaces obtained as differences between brick matchings.

\paragraph{Example 1: Phase boundaries.}

{\it Phase boundaries} are $2d$ surfaces on a brane brick model, which are in one-to-one correspondence with edges of the toric diagram of the corresponding CY$_4$. More precisely, the homology of such a surface on $\mathbb{T}^3$ is equal to the $\mathbb{Z}^3$ vector defining the corresponding edge. Phase boundaries are the brane brick model analogues of zig-zag paths for brane tilings. 

We refer to the corners of toric diagrams as {\it extremal} points. Let us consider two extremal brick matchings $p_\mu$ and $p_\nu$, with coordinates 
\beq
\begin{array}{ccc}
p_\mu: & \ \ & (m_x,m_y,m_z) \\ 
p_\nu: & \ \ & (n_x,n_y,n_z)
\end{array}
\eeq
and connected by an edge of the toric diagram. The phase boundary associated to the edge connecting them is $\eta_{\mu\nu}=p_\mu-p_\nu$, and its homology on $\mathbb{T}^3$ is $(m_x-n_x,m_y-n_y,m_z-n_z)$.\footnote{More generally, if the edge between $p_\mu$ and $p_\nu$ consists of $n$ segments, $p_\mu-p_\nu$ gives rise to $n$ disconnected surfaces on the brane brick model, i.e. $n$ phase boundaries, with the same homology.} 

\begin{figure}[ht!]
	\centering
	\includegraphics[height=6cm]{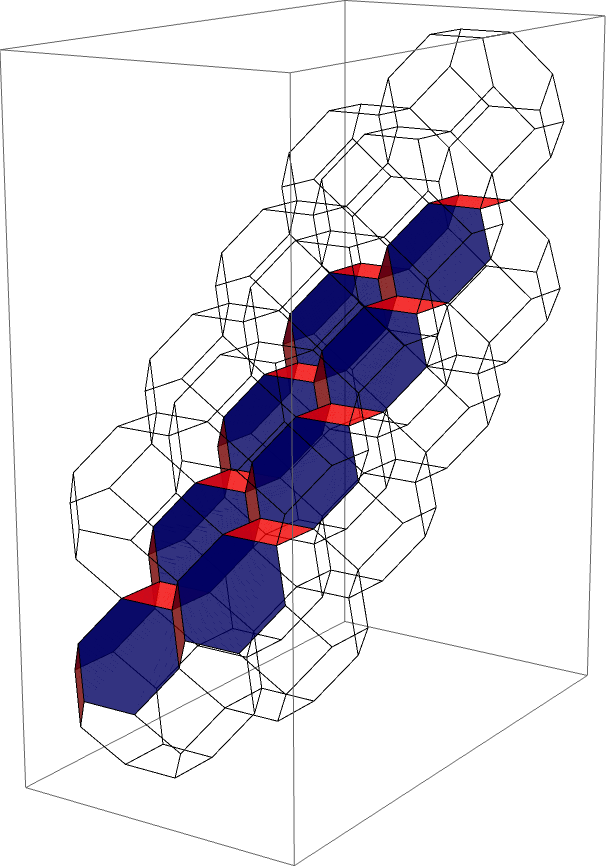}
\caption{Phase boundary $\eta_{yw}=p_y-p_w$, calculated using the perfect matchings in Figures \ref{pms_C4} and \ref{toric_C4_with_pms} (including Fermis).}
	\label{phase_boundary_C4}
\end{figure}

In the same way that zig-zag paths of brane tilings are in one-to-one correspondence with external legs of the $(p,q)$ web dual to the toric diagram of the corresponding CY$_3$, phase boundaries are in one-to-one correspondence with $2d$ ``legs" of $(p,q,r)$-webs dual to the $3d$ toric diagram of the corresponding  CY$_4$. \fref{phase_boundary_C4} shows a phase boundary for $\mathbb{C}^4$, represented on the universal cover of the brane brick model.

Once we introduce the crystal melting model for toric CY 4-folds in the coming sections, we will see that, very much like their zig-zag cousins, phase boundaries are associated to the interfaces between regions of the crystal with different asymptotic behavior.

\paragraph{Example 2: Difference with respect to a reference brick matching.}

The difference between a brick matching and a reference brick matching is used to determine the corresponding height function. As we will explain in Section \sref{section_height_function}, the height function jumps by 1 at each of the resulting surfaces. An explicit example will be presented in Section \sref{section_melting_configurations_BBM}.

\section{Height function}

\label{section_height_function}

Given a brick matching $p_\mu$ of either a brane brick model (i.e. with periodic identifications on $\mathbb{T}^3$) or its universal cover, it is possible to define an integer-valued {\it height function} $h_\mu$. To do so, we first pick a reference brick matching $p_0$. As discussed above, the difference $p_\mu-p_0$ defines a set of oriented surfaces. The height function jumps by $\pm 1$ when traversing these surfaces, with the sign determined by the orientation of the crossing. This prescription only determines changes of the height function so, in order to fully fix it it, it is necessary to specify its value at some point of the brane brick model.

When considering brane brick models, which live on $\mathbb{T}^3$, instead of their universal cover, the height function is not singled valued. In this case, it is more appropriate to consider the {\it slope of the height function}. Is is defined as $(\Delta_1 h_\mu, \Delta_2 h_\mu, \Delta_3 h_\mu)$, where $\Delta_i h_\mu$ is the change in the height function when going between consecutive copies of the unit cell along the $i=1,2,3$ fundamental direction of the torus.

\section{A statistical model of crystal melting for toric CY$_4$}

\label{section_model_crystal_melting_CY4_quiver}

In this section, we introduce the statistical model of crystal melting associated to the D0-D8 system in $\mathbb{C}^4$. This theory is described by the flavored quiver in \fref{quiver_C4_flavor}.\footnote{In the literature, when discussing analogous constructions for CY 3-folds, the terms {\it crystal} and {\it atoms} are sometimes replaced by {\it pyramids} and {\it stones}, respectively.} The starting point for constructing the crystal is the universal cover of the periodic quiver for $\mathbb{C}^4$, which we denote $\tilde{Q}$. 

We denote the $3d$ space in which $\tilde{Q}$ lives as {\it quiver space} and assign coordinates $(x,y,z)$ to it.\footnote{These coordinates will be useful for constructing $\tilde{Q}$ and the crystal later. In this paper we do not assign a physical meaning to their actual values, beyond determining the structure of the periodic quiver.} The flavor arrow $q$, often referred to as {\it framing} arrow in this context, is assigned to one of the nodes in $\tilde{Q}$, effectively determining the origin in quiver space.

The model is a natural generalization to toric CY 4-folds of the one for toric CY 3-folds introduced in \cite{Ooguri:2009ijd} (see also \cite{Chuang:2009crq,Eager:2011ns,Nishinaka:2013mba,Bao:2022oyn} for relevant discussions and generalizations).

\subsection{The unmolten crystal}

\label{section_statistical_model_unmolten}

Without loss of generality, let us assume that the flavor $q$ is connected to node $i_0$ in $\tilde{Q}$. We define the crystal such that every atom in it is in one-to-one correspondence with an oriented path of chiral fields in $\tilde{Q}$ starting from $q$ (equivalently starting from $i_0$), modulo $J$- and $E$-term relations.\footnote{$J$- and $E$-term relations lead to equivalences between paths in the periodic quiver with the same endpoints. In other words, atoms can generically be reached in multiple, equivalent ways.} The crystal is built out of atoms stacked on top of the nodes of $\tilde{Q}$ on $\mathbb{R}^{3}$. Since the quiver for $\mathbb{C}^4$ has a single node, the crystal has a single type of atom. The crystal contains a fourth dimension, that we will denote the {\it depth} $d$.Chiral fields determine the relative depth of the atoms connected by them. If there is a chiral arrow from atom $i$ to atom $j$, atom $j$ is at a higher depth than $i$. We can think about two such atoms as partially overlapping.\footnote{We reserve the term {\it overlapping} for atoms that are directly on top of each other.}

To build the crystal, we first place an atom over node $i_0$, which will become the tip of the crystal. Then, we iteratively add new atoms according to the chiral fields that emanate from the corresponding nodes in the original quiver. The $(x,y,z)$ position of an atom is the one of the corresponding node in $\tilde{Q}$, while the depth is  proportional to the $R$-charge (equivalently the conformal dimension) of the corresponding chiral operator. In the simple case of $\mathbb{C}^4$, in which the four chiral fields are equivalent, this is simply proportional to the length of the path (namely the number of chirals in it).\footnote{The explicit values of the depth are not necessary for determining melting configurations, as we discuss in the coming sections.} This procedure results in an infinite crystal.

Modulo $J$- and $E$-term constraints, every oriented path $\gamma_{i_0,j}$ defining atom $j$ of the crystal can be expressed as 
\beq
\gamma_{i_0,j}=v_{i_0,j} \omega^n
\eeq
where $v_{i_0,j}$ is a shortest path connecting $i_0$ to $j$, $\omega$ is the closed loop associated to a chiral cycle, and $n\geq 0$.\footnote{As we explained in Section \sref{section_brick_matchings}, all chiral cycles are equivalent up to $J$- and $E$-terms, and therefore can be identified with a single variable $\omega$.} We can interpret $v_{i_0,j}$ as defining an atom at the top layer of the crystal.  An atom with an additional factor of $\omega^n$, is located directly below, $n$ levels down.

\subsection{Melting configurations}

\label{section_melting_configurations}

We now consider consider {\it molten crystals}, i.e. configurations that are obtained by removing atoms from the unmolten crystal. We will denote any crystal configuration (i.e. molten or not) as $\mathcal{I}_\mu$. We define the corresponding complement $\Omega_\mu$ as the difference between the unmolten crystal and $\mathcal{I}_\mu$, i.e. $\Omega_\mu$ is the set of removed atoms. For brevity, we will refer to the $\Omega_\mu$ as {\it melting configurations}.\footnote{Anticipating the connection between molten crystals and brick matchings that will be discussed in Section \sref{section_crystal_melting_BBM}, we use the same type of subindex to label both.}

Let us momentarily focus on the unflavored quiver $Q$. Let us denote $Q_0$ and $Q_X$ the sets of nodes and chiral arrows in $Q$, respectively. The set of all open oriented chiral paths in $Q$ gives rise to an algebra $\mathbb{C}[Q_0,Q_X]$, that we will call the {\it chiral path algebra}. Given the ideal of relations coming from vanishing $J$- and $E$-terms
\beq
\mathcal{I}_{J, E} = \langle J_{a}^{+} - J_{a}^{-} = 0, E_{a}^{+} - E_{a}^{-} = 0 \rangle \, ,
\eeq
where $a$ runs over all Fermis, it is natural to define the factor algebra $A=\mathbb{C}[Q_0,Q_X]/ \mathcal{I}_{J, E}$. $A$ consists of the open chiral paths in the unflavored quiver modulo vanishing $J$- and $E$-terms.

Melting configurations are constructed according to the following {\it melting rule}.

\medskip

\begin{center}
\begin{tabular}{| m{0.85\textwidth}|  }
\hline 
{\bf Melting rule:}  If $\gamma_{i_0,i} \alpha_{i,j}$ is in $\Omega_\mu$ for some $\alpha_{i_j} \in A$, then $\gamma_{i_0,i}$ should also be in $\Omega_\mu$. 
\\ \hline
\end{tabular}
\end{center}

\medskip

\noindent Heuristically, this means that if an atom is removed in a given melting configuration, then all atoms on top of it must be removed too. More precisely, starting from atom $j$, we can go up the crystal by following the path $\alpha_{i,j}$ in the reverse direction, encountering atom $i$, which should also be removed. 

It is straightforward to see that every molten crystal $\mathcal{I}_\mu$ defines an ideal of $A$. To show this, we consider the contraposition of the melting rule, which implies that for any $\gamma \in \mathcal{I}_\mu$ and any $\alpha \in A$, then $\gamma \alpha$ is also in $\mathcal{I}_\mu$. In simple words, starting from any atom in a molten crystal and moving from it along a path $\alpha \in A$, results in another atom in the molten crystal, i.e. an atom that has not been removed.

Explicit constructions of the unmolten crystal and melting configurations will be presented in Sections \sref{section_model_crystal_melting_CY4_quiver} and \sref{section_exploring_the_crystal}.

\paragraph{Melting height.}

We have previously introduced the notion of depth, which measures how far below the tip of the unmolten crystal an atom is. When studying melting configurations, it is useful to consider the {\it melting height} $h$, which is a also a function of the point $(x,y,z)$ in quiver space. For a crystal configuration $\mathcal{I}_\mu$, it is defined as
\beq
h=d_0-d_\mu \, ,
\eeq
where $d_0$ and $d_\mu$ are the depth functions of the top layers of the unmolten crystal and $\mathcal{I}_\mu$, respectively. It is possible to normalize the depth such that $h$ counts the number of removed atoms in the melting configuration $\Omega_\mu$ at every value of $(x,y,z)$. From now on, we will assume such normalization for the melting height.

\paragraph{Partition functions.}
It is useful to define a partition function of the form
\beq
Z=\sum_{\Omega_\mu} y^{n_\mu} \, ,
\eeq
where $n_{\mu}$ is the number of atoms in melting configuration $\Omega_\mu$. The integer coefficient of $Z$ at order $y^n$ therefore gives the number of melting configurations with $n$ atoms.

In terms of branes, the unmolten crystal represents the single D8-brane, while every melting configuration corresponds to adding $n_\mu$ units of D0-charge.

\section{The $4d$ crystal for $\mathbb{C}^4$}

\label{section_crystal_C4}

In this section, we construct the unmolten crystal for $\mathbb{C}^4$. According to the periodic quiver in \fref{quiver_BBM_C4}, we assign the following vectors to each type of chiral field
\beq
\begin{array}{ccl}
v_X & = & (1,-1,1) \\[.1 cm]
v_Y & = & (1,1,-1) \\[.1 cm]
v_Z & = & (-1,1,1) \\[.1 cm]
v_W & = & (-1,-1,-1) 
\end{array}
\label{generating_vectors_quiver}
\eeq
The coordinates of every atom in quiver space are then given by
\beq
(x,y,z)=n_X v_X + n_Y v_Y + n_Z v_Z + n_W v_W  
\eeq
with $n_X, n_Y, n_Z, n_W\geq 0$ the numbers of $X$, $Y$, $Z$ and $W$ fields in a path connecting the origin to the atom under consideration. The fourth coordinate of every atom is the depth, which is given by\footnote{For simplicity, we normalize the depth such that it is simply given by the number of chiral fields in the shortest path connecting an atom to the origin.}
\beq
d = n_X + n_X + n_Z + n_W \,.
\eeq  
An atom might be reached by different paths, due to the equivalences coming from vanishing $J$- amd $E$-terms. Therefore, a given atom might be associated to different values of $(n_X,n_Y,n_Z,n_W)$.

\subsection{Slicing $4d$ crystals}

\label{section_slicing_crystals}

\begin{figure}[ht!]
	\centering
	\includegraphics[width=\textwidth]{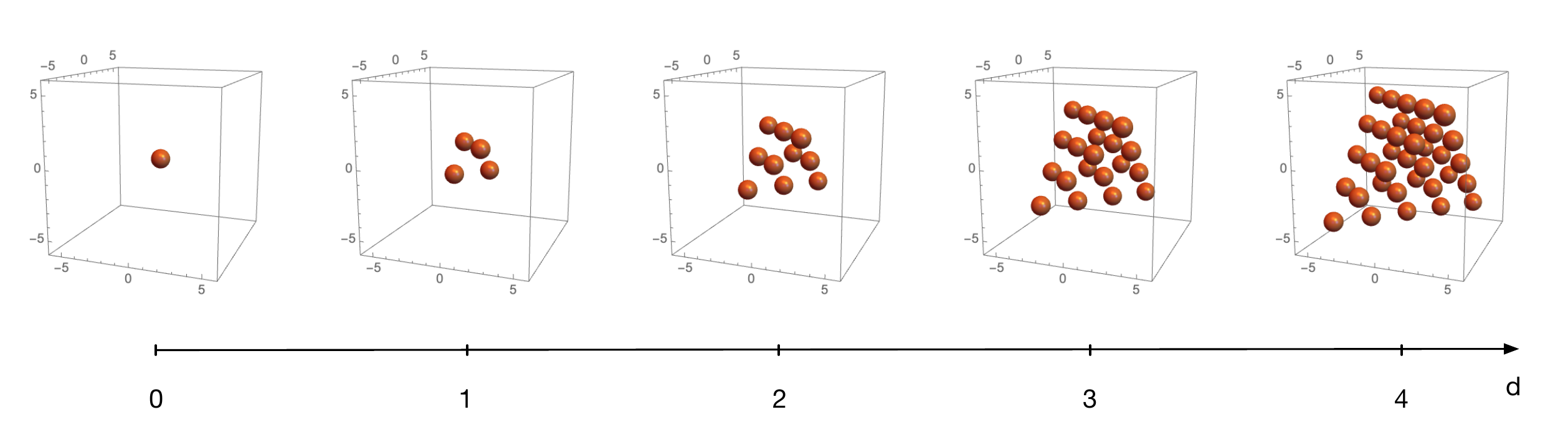}
\caption{Constant depth slices of the unmolten crystal for $\mathbb{C}^4$ up to $d=4$.}
	\label{C4_stones_vs_depth}
\end{figure}

Visualizing a $4d$ crystal is of course challenging. In this paper we will present various ways to do so. The first approach consists of {\it slicing} the crystal at different depths. This approach can be applied to both unmolten and molten crystals. Let us illustrate this construction with the unmolten crystal. \fref{C4_stones_vs_depth} shows the atoms in it up to $d=4$.

From \fref{C4_stones_vs_depth}, we see that the $3d$ slice at depth $d$ is a tetrahedron of edge length $d+1$. The number of atoms at depth $d$ is therefore 
\beq
n(d)={1\over 6} (1+d) (2+d) (3+d) \, ,
\label{n_stones_d}
\eeq
which results in
\beq
\begin{array}{|c|c|c|c|c|c|c|c|c|c|c|c|}
\hline 
d & 0 & 1 & 2 & 3 & 4 & 5 & 6 & 7 & 8 &  9 & 10 \\ \hline
n(d) & \ \, 1 \, \ \ & \ \, 4 \, \ \ & \ 10 \ \ & \ 20 \ \ & \ 35 \ \ & \ 56 \ \ & \ 84 \ \ & 120 & 165 & 220 & 286 \\ \hline
\end{array}
\eeq

Equation \eref{n_stones_d} can be generalized to $D$ dimensions, for which the number of atoms in the $(D-1)$-dimensional slices at depth $d$ is
\beq
n_D(d)={1\over (D-1)!} (1+d) (2+d) ... (D-1+d) \, .
\eeq

\smallskip

\paragraph{Overlapping atoms.}
The first repeated $3d$ position $(x,y,z)$, i.e. the first case of an atom with another one directly on top, occurs at depth 4, where we encounter a second atom at $(0,0,0)$. At depth 5, we get 4 repeated atoms at positions $(-1, -1, -1)$, $(-1, 1, 1)$, $(1, 1, -1)$ and $(1, -1, 1)$. More generally, at depth $d$, we get atoms with the same positions in quiver space as all the ones at depth $d-4$.

\smallskip

\paragraph{The crystal and toric geometry.}
The tetrahedral shape of the slices is related to the tetrahedral shape of the toric diagram, which is shown in \fref{toric_C4}. Each of the vertices in a slice corresponds to a vertex in the toric diagram, a fact that can be understood as follows. At a fixed depth $d = n_X + n_X + n_Z + n_W$, a vertex corresponds to a direction along which the distance from the origin is maximized. This corresponds to using $d$ copies of the same vector $v_i$, $i=X,Y,Z,W$. Also, edges between two faces in a slice corresponds to edges between the two corresponding vertices in the toric diagram, and so on. We will revisit the correspondence between the crystal and the underlying toric geometry in Section \sref{section_crystal_melting_BBM}. This connection will become even more tangible in the reformulation of the model in terms of brane brick models.

\subsection{The ``$3d$ surface" of the $4d$ crystal}

Let us definite the ``$3d$ surface” of a $4d$ crystal configuration $\mathcal{I}_\mu$ as its top layer, in analogy with the $2d$ surface of an ordinary $3d$ crystal.\footnote{We feel it is useful to stick to the term surface to denote the top layer of a crystal due to its intuitive interpretation, despite that, in this case, it is a 3-dimensional object.} In other words, we define:

\bigskip

\noindent {\bf Surface of the crystal $S_\mu$:} set of all the atoms in $\mathcal{I}_\mu$ such that their depth is minimum for a given position in quiver space. The $4d$ coordinates of the atoms are $(x,y,z,d)$. 

\bigskip

A corollary of this definition is that if we consider all chiral arrows connecting atoms in $S_\mu$, they are such that they do not form closed oriented loops. If there was such a loop, we would have two atoms at the same $(x,y,z)$ position but at different values of $d$. Therefore, one of them could not be in $S_\mu$. We will revisit this fact in Section \sref{section_crystal_melting_BBM} where we present a reformulation of the crystal in terms of brane brick models.
 
\begin{figure}[ht!]
	\centering
	\includegraphics[width=\textwidth]{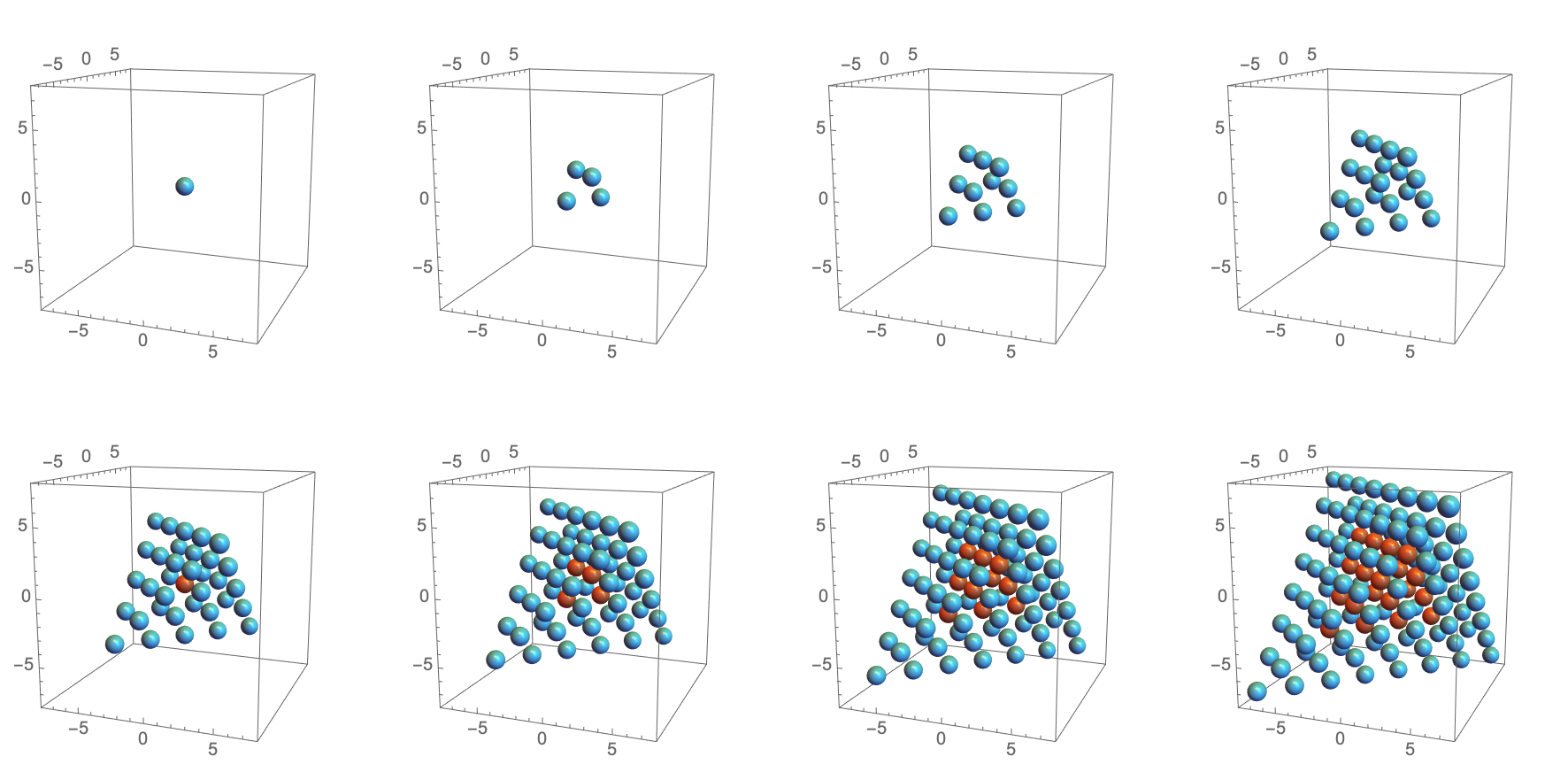}
\caption{Atoms on the surface of the unmolten crystal for $d=0,\ldots, 7$.}
	\label{C4_stones_surface}
\end{figure}

As an example, let us consider the surface of the unmolten crystal. Algorithmically, it can be constructed following the same procedure we used to build the full crystal, but keeping only the atoms of lowest depth for every $(x,y,z)$ coordinate. With the normalization we are using, the depths of overlapping atoms differ by multiples of 4, so at every depth $d$, we need to remove the atoms at every $d'<d$ such that $d-d'=0 \mod 4$. \fref{C4_stones_surface} shows the atoms on the surface up to $d=7$. Atoms on the surface and the interior of the crystal are shown in blue and orange, respectively. We observe that up to $d=3$ all atoms are on the surface.

\subsection{Melting configurations and solid partitions}

\label{section_crystal_and_solid_partitions}

The crystal for $\mathbb{C}^4$ is equivalent to a $4d$ corner, with melting configurations in one-to-one correspondence with solid partitions. The connection between the quiver theory for D0 and D8-branes on $\mathbb{C}^4$ and solid partitions was first explored in the context of the {\it Magnificent Four} model \cite{Nekrasov:2017cih,Nekrasov:2018xsb,Nekrasov:2023nai}. The partition function counting solid partitions is
\beq
Z=1+y+4 y^2+10 y^3+26 y^4+59 y^5+140 y^6 + \cdots
\label{Z_solid_partitions}
\eeq
In the coming section, we will illustrate how this partition function arises from the crystal model.

\section{Exploring the crystal}

\label{section_exploring_the_crystal}

\begin{figure}[ht!]
	\centering
	\includegraphics[height=9cm]{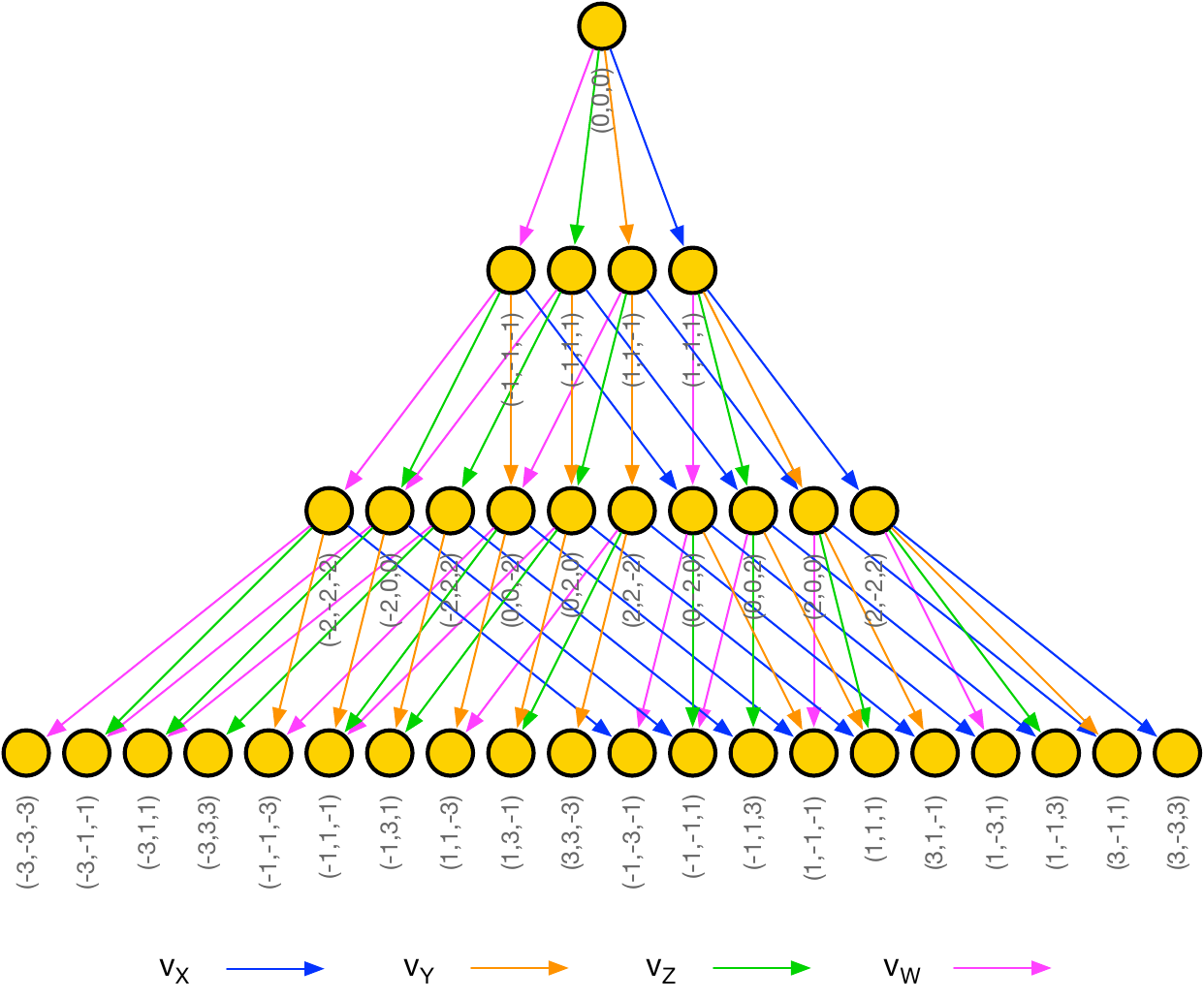}
\caption{Hasse diagram for the $\mathbb{C}^4$ crystal up to $d=3$.}
	\label{poset_C4}
\end{figure}

In this section we present additional tools for visualizing and studying $4d$ crystals. All the atoms in the crystal define a poset $\Delta$, in which the ordering is determined by the partial overlap relations. An efficient way of keeping track of the latter is by means of a {\it Hasse diagram}. An arrow in this diagram from atom $a$ to atom $b$ indicates that $a$ is on top of $b$. In this context, the arrows actually correspond to chiral fields in the quiver. \fref{poset_C4} shows the poset for the unmolten crystal for $\mathbb{C}^4$ up to $d=3$.

The crystal under consideration has a single top atom, and hence a single atom at the peak of the Hasse diagram. More general crystals, like the ones described in Section \sref{section_crystal_general_CY4}, might have multiple top atoms, which would be reflected in the respective Hasse diagrams. Each layer in \fref{poset_C4} corresponds to one of the slices in Section \sref{section_slicing_crystals}. While it is natural to vertically organize the Hasse diagram according to the depth, only the relational structure stemming from the arrows is important.

\subsection{Melting configurations}

\label{section_pyramid_partitions}

Here we present explicit examples illustrating how the Hasse diagram can be used to classify melting configurations. According to the melting rule, whenever an atom is removed from the crystal, all atoms above it, i.e. all atoms contained in downward paths terminating in it, should also be removed.

\begin{figure}[ht!]
	\centering
	\includegraphics[height=5cm]{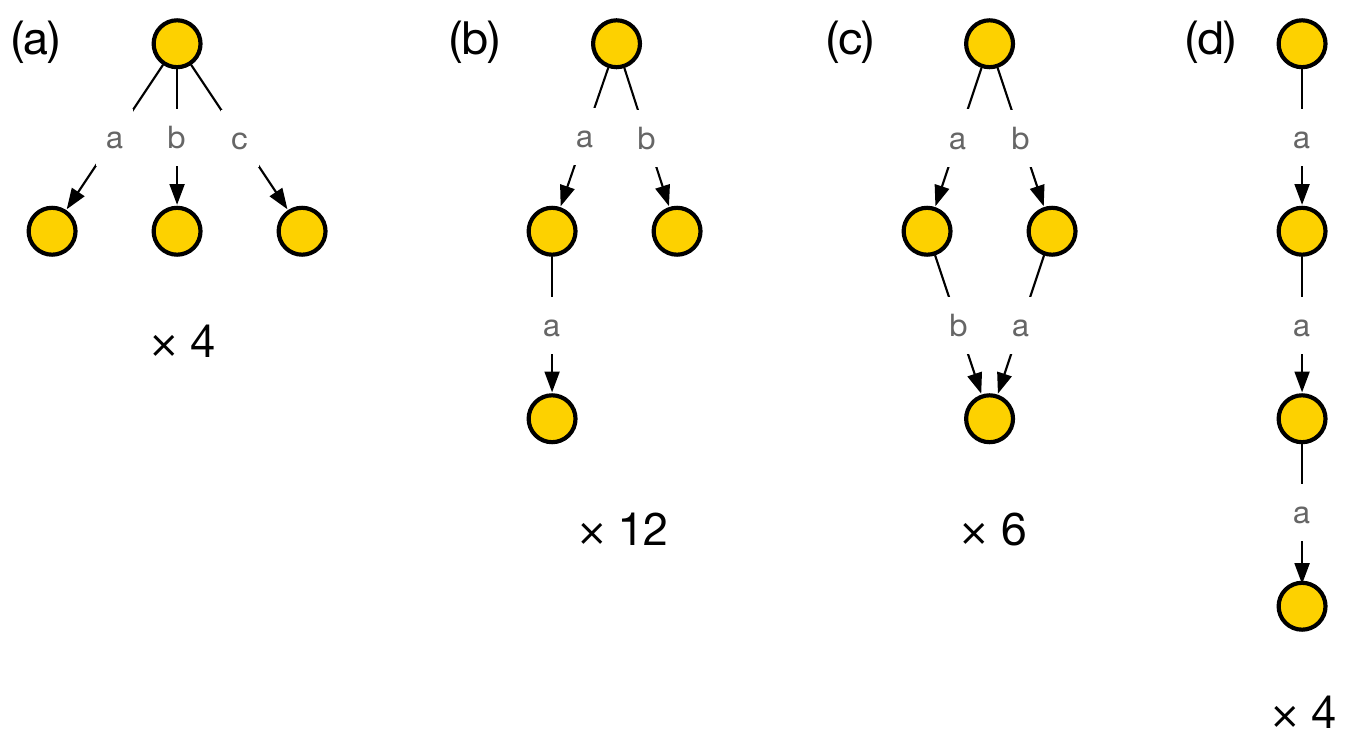}
\caption{Hasse diagram representations of all melting configurations with 4 atoms.}
	\label{configurations_4_stones}
\end{figure}

\paragraph{Example 1: Melting configurations with 4 atoms.}
Let us count the number of melting configurations with 4 atoms, or equivalently, according to the discussion in Section \sref{section_crystal_and_solid_partitions}, the number of solid partitions with 4 boxes. The Hasse diagram provides an efficient way to represent and count all such configurations. They are given in \fref{configurations_4_stones}, together with their multiplicities, which are easily determined by the combinatorics of the types of fields involved in each type of configuration. Types of fields are indicated with gray letters over the arrows. We conclude that there are 26 melting configurations with 4 atoms, which agrees with the corresponding term in the partition function \eref{Z_solid_partitions}.

\begin{figure}[ht!]
	\centering
	\includegraphics[height=11cm]{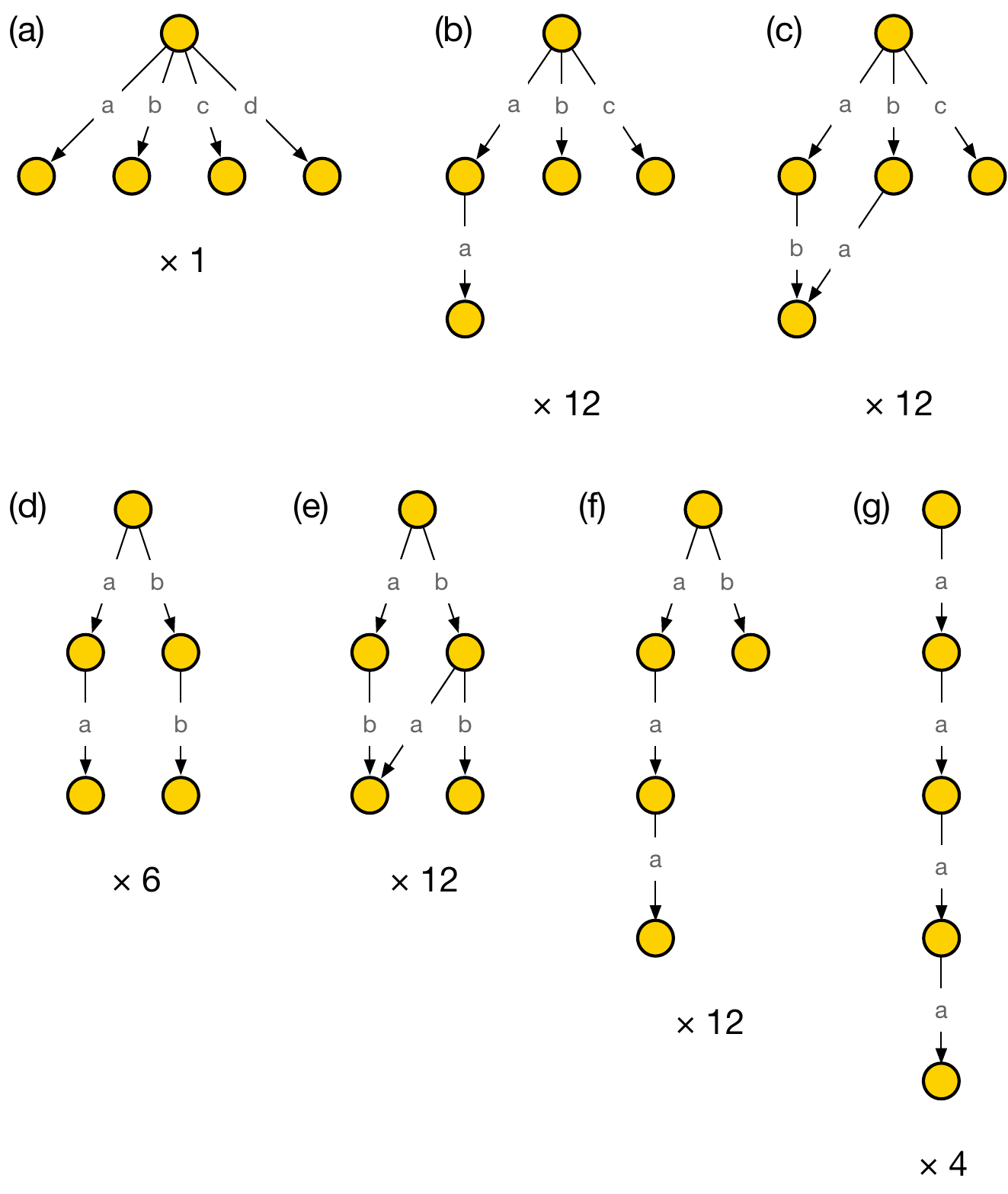}
\caption{Hasse diagram representations of all melting configurations with 5 atoms.}
	\label{configurations_5_stones}
\end{figure}

\paragraph{Example 2: Melting configurations with 5 atoms.}
We can perform a similar exercise and use the Hasse diagram to count the number of melting configurations with 5 atoms. They are shown in \fref{configurations_5_stones} with their multiplicities. We conclude that there are 59 melting configurations with 5 atoms, in agreement with \eref{Z_solid_partitions}.

\paragraph{Example 3: A melting configuration with overlapping atoms.}

\begin{figure}[ht!]
	\centering
	\includegraphics[height=8cm]{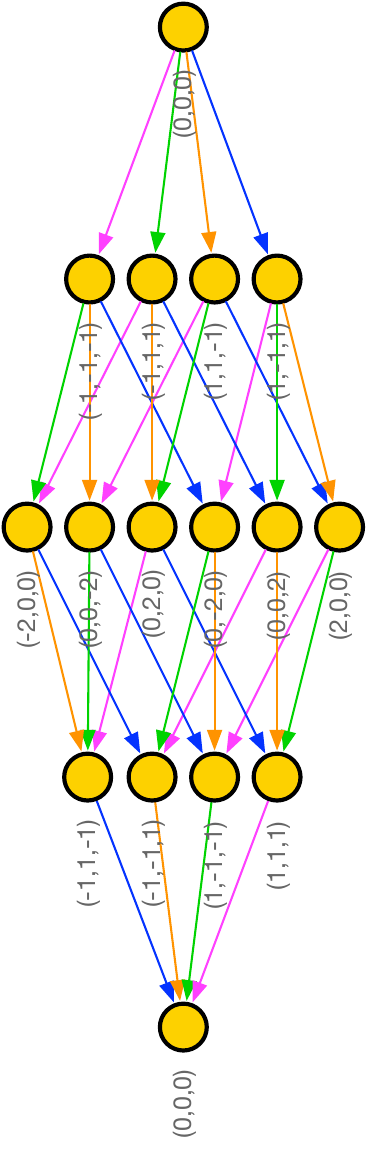}
\caption{Subset of the Hasse diagram for $\mathbb{C}^4$ representing the minimal melting configuration containing the first pair of overlapping atoms.}
\label{poset_height_2_configuration}
\end{figure}

Let us use the Hasse diagram to identify the smallest melting configuration containing a point $(x,y,z)$ with melting height greater than 1, i.e. the smallest one containing at least two overlapping atoms. As mentioned in Section \sref{section_slicing_crystals}, the first overlapping atom is located at $(x,y,z,d)=(0,0,0,4)$, so let us look for the minimal melting configuration containing it. Our previous discussion implies that this melting configuration corresponds to the subset of the poset in \fref{poset_C4} that consists of all atoms contained in downward paths terminating in $(0,0,0,4)$. Equivalently, we start from the corresponding point in the Hasse diagram and move upwards, removing all atoms connected to it by arrows and iterating this process starting from the newly deleted atoms. The resulting subset of the Hasse diagram is shown in \fref{poset_height_2_configuration}, from where we see that this melting configuration contains 16 atoms. Larger melting configurations containing the $(0,0,0,4)$ atom must include this subset.

It is interesting to note that this melting configuration, which is the first one with an atom at melting height 2, appears at a relative high order in the partition function, $q^{16}$. As it follows from our analysis above, this follows from the melting rule. Heuristically, the higher the dimension of the crystal, the larger the number of ways in which atoms can partially overlap. Our result also implies that all melting configurations with 15 or fewer atoms are effectively 3-dimensional.

\begin{figure}[ht!]
	\centering
	\includegraphics[width=\textwidth]{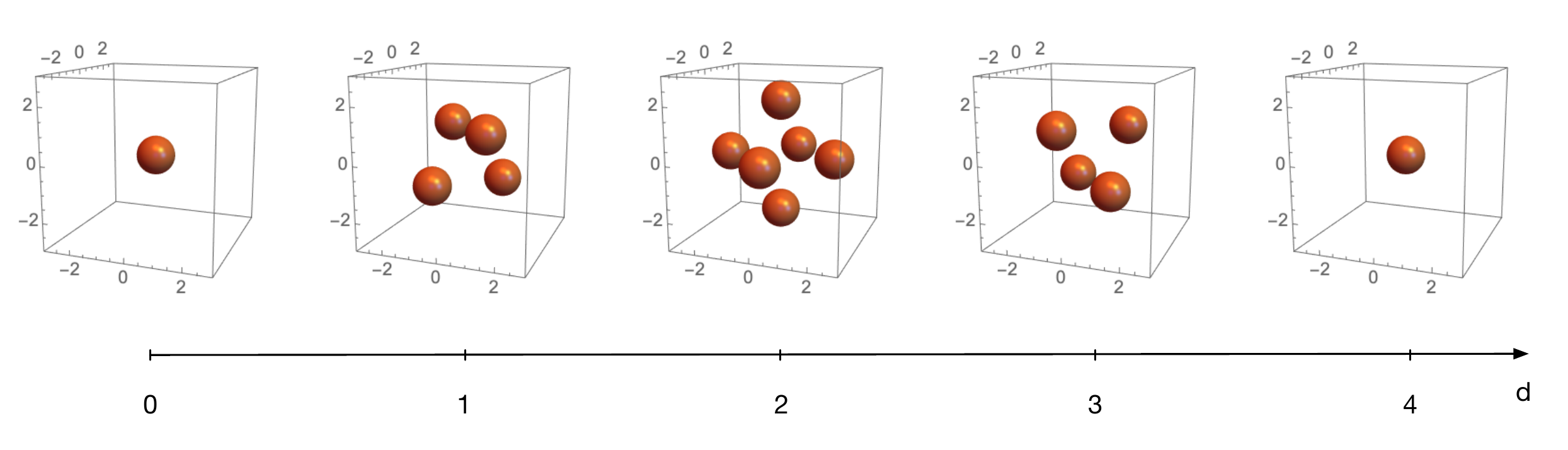}
\caption{Constant depth slices of the melting configuration defined by \fref{poset_height_2_configuration}.}
	\label{C4_16_stones_sliced}
\end{figure}

A better idea of how the structure of this melting configuration is obtained by considering its slices according to depth, as shown in \fref{C4_16_stones_sliced}.

\begin{figure}[ht!]
	\centering
	\includegraphics[height=4cm]{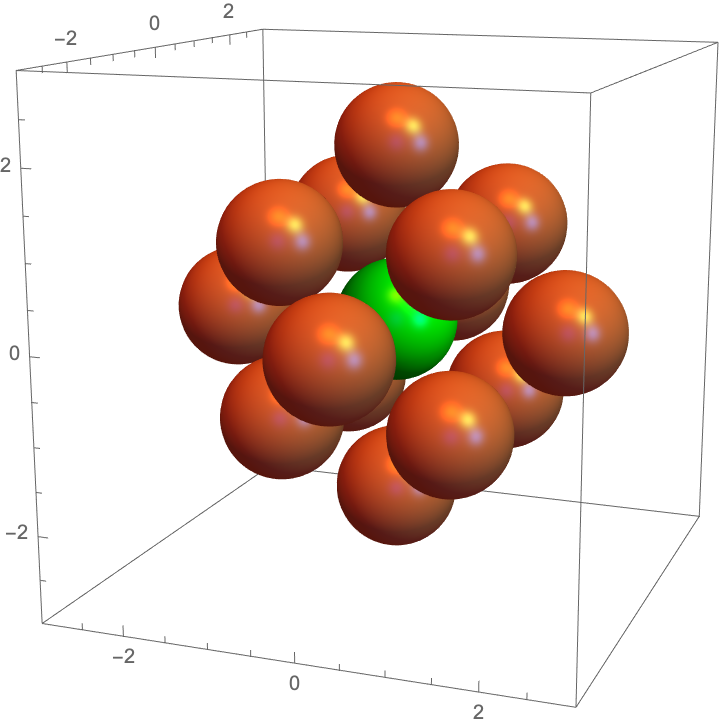}
\caption{Melting configuration given by \fref{poset_height_2_configuration}. Orange and green spheres correspond to melting height equal to 1 and 2, respectively.}
	\label{C4_16_stones_together}
\end{figure}

An alternative visualization is achieved by projecting the configuration onto the $(x,y,z)$ space and including information regarding the melting height (not to be confused with the depth used in \fref{C4_16_stones_sliced}). This is done in \fref{C4_16_stones_together}.

\section{An appetizer: brane tilings and crystal melting for $\mathbb{C}^3$}

\label{section_appetizer}

\begin{figure}[ht!]
	\centering
	\includegraphics[height=3.5cm]{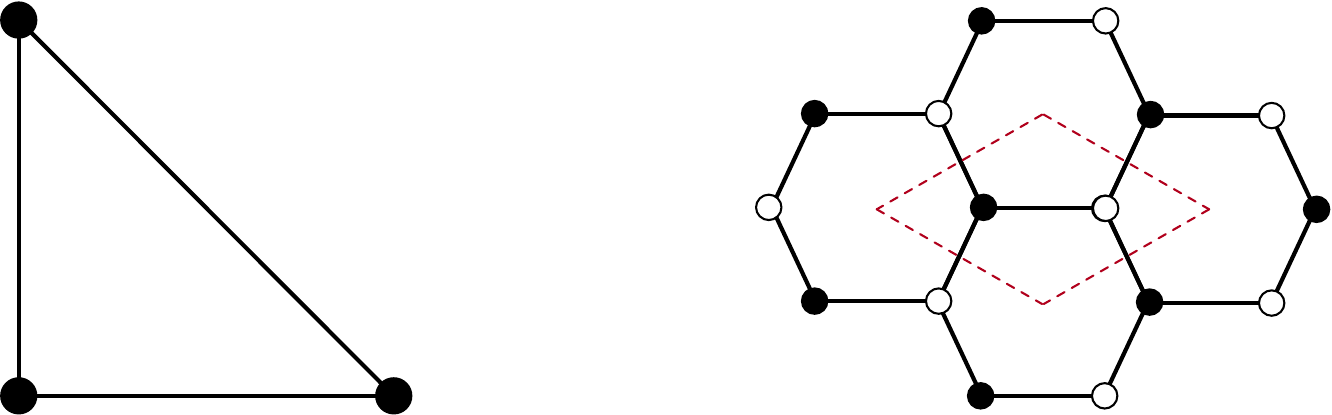}
\caption{Toric diagram for $\mathbb{C}^3$ and the corresponding brane tiling. Dashed red lines indicate the boundaries of the unit cell.}
	\label{toric_brane_tiling_C3}
\end{figure}

In Section \sref{section_crystal_melting_BBM}, we will introduce a brane brick model description of the crystal melting model studied above. Given the challenges of visualizing $4d$ objects, it is useful to first review the analogous construction in $3d$, i.e. the brane tiling and crystal melting model for D0-branes and a D6-brane on $\mathbb{C}^3$. We refer the reader to \cite{Franco:2005rj} for background on brane tilings and \cite{Ooguri:2009ijd} for further details of the ideas in this section. The gauge theory on D3-branes probing $\mathbb{C}^3$ is $4d$ $\mathcal{N}=4$ SYM. As we did before, we will interpret this theory as a quiver quantum mechanics on D0-branes. The D6-brane adds a chiral flavor going from a global node representing the D6-brane into the only node of the quiver. \fref{toric_brane_tiling_C3} shows the toric diagram for $\mathbb{C}^3$ and the corresponding brane tiling, which is the hexagonal lattice.\footnote{Larger unit cells on the hexagonal lattice correspond to Abelian orbifolds of $\mathbb{C}^3$ \cite{Hanany:2005ve,Franco:2005rj,Hanany:2010cx,Davey:2010px,Davey:2011dd}.}

This simple theory has three perfect matchings, which are in one-to-one correspondence with the chiral fields associated to the three complex directions. The $P$-matrix summarizing the perfect matchings is therefore
\beq
P=\left(\begin{array}{c|ccc} 
& \ \ p_x \ \ & \ \ p_y \ \ & \ \ p_z \ \ \\ \hline
X & 1 & 0 & 0 \\
Y & 0 & 1 & 0 \\
Z & 0 & 0 & 1 
\end{array}\right) \, .
\eeq

\begin{figure}[ht!]
	\centering
	\includegraphics[height=4cm]{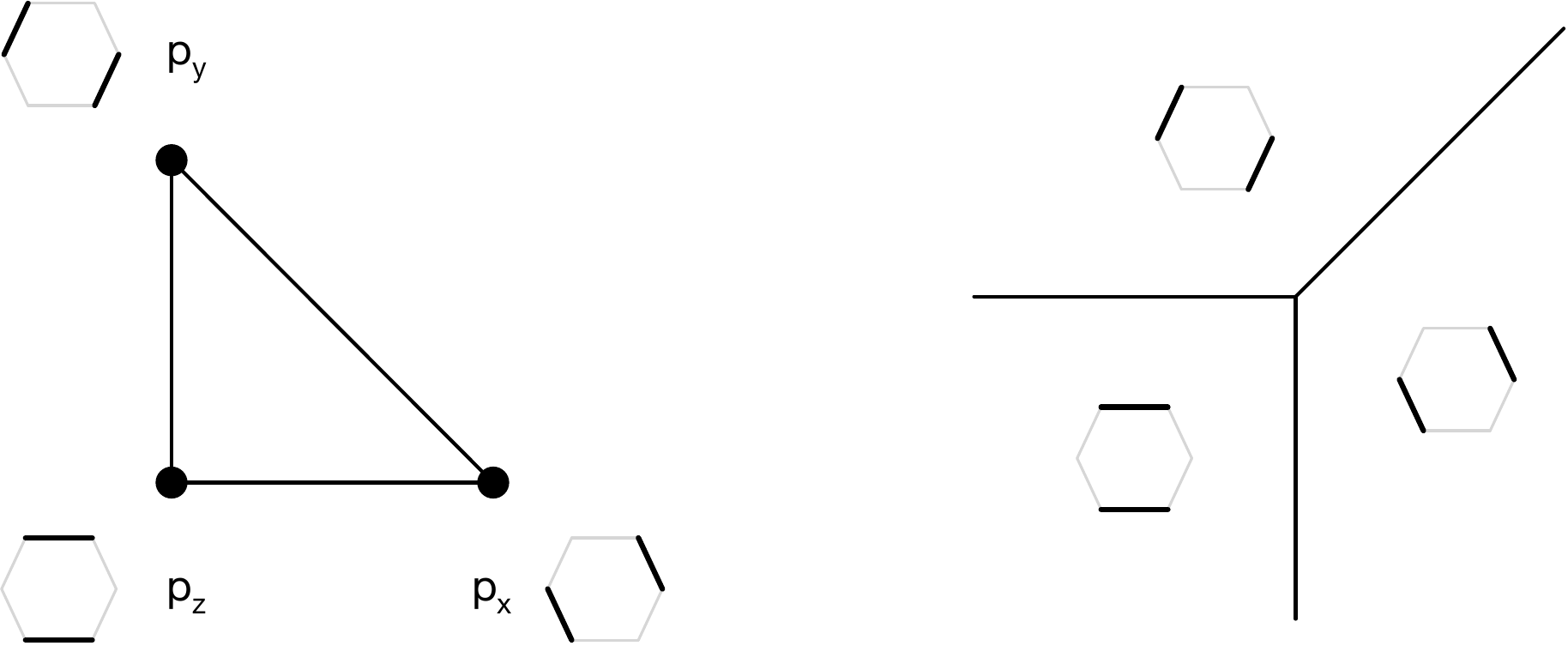}
\caption{Correspondence between perfect matchings for $\mathbb{C}^4$ and points in its toric diagram. On the right, we show the dual $(p,q)$ web.}
	\label{toric_C3_with_pms}
\end{figure}

\fref{toric_C3_with_pms} shows the correspondence between these perfect matchings and points in the toric diagram, in analogy with \fref{toric_C4_with_pms}. It also shows the dual $(p,q)$ web, in which perfect matchings map to regions separated by the lines in the web.

The underlying structure of the crystal melting model is the universal cover of the brane tiling for $\mathbb{C}^3$, namely an infinite hexagonal lattice.\footnote{While we will focus on the brane tiling perspective, the crystal melting model can also be formulated in terms of the universal cover of the periodic quiver.} Melting configurations are in one-to-one correspondence with perfect matching of this extended tiling, as we explain below.

\subsection{The unmolten crystal}

\label{section_unmolten_crystal_C3}

\begin{figure}[ht!]
	\centering
	\includegraphics[height=5cm]{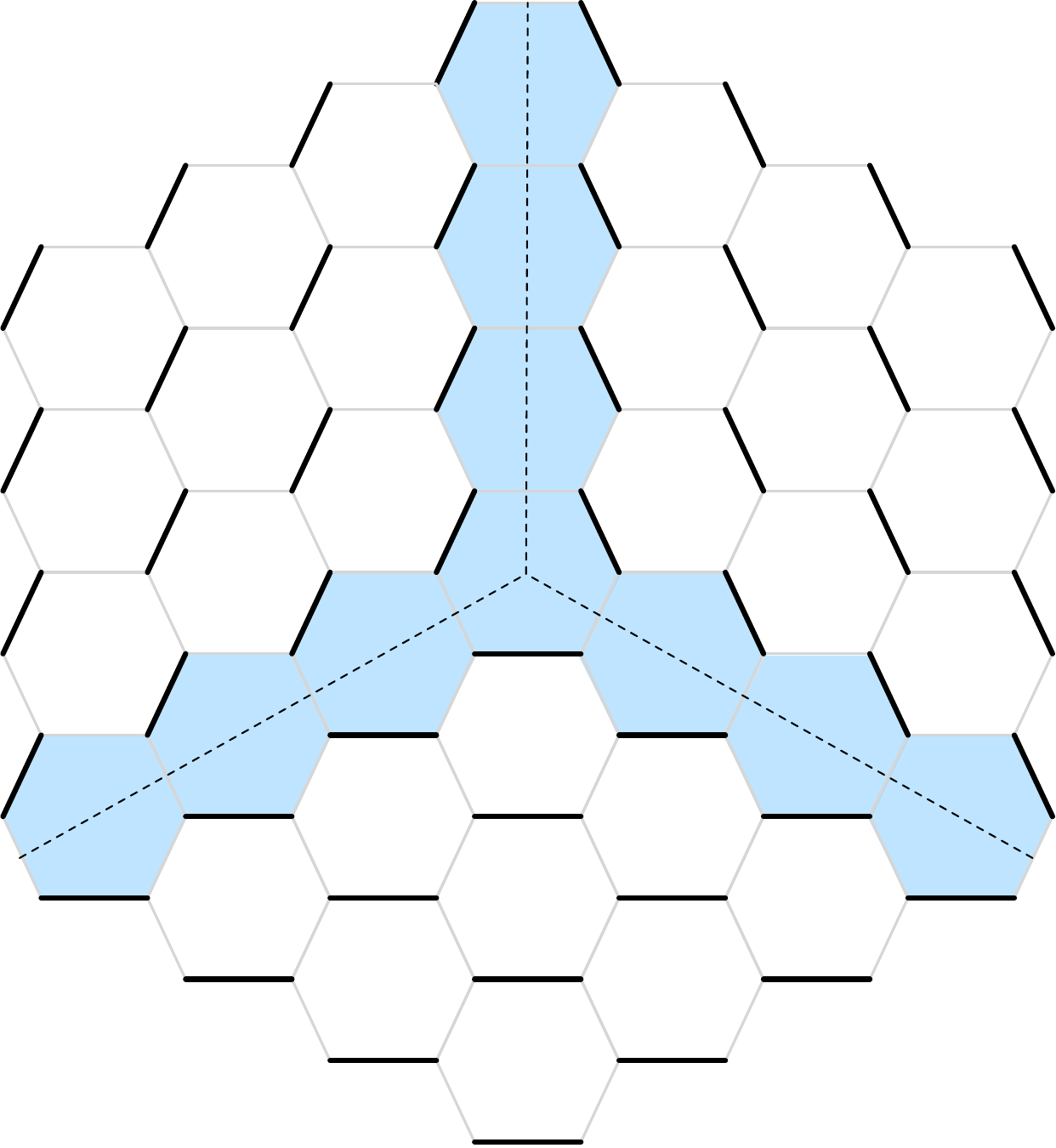}
\caption{The canonical perfect matching for $\mathbb{C}^3$.}
	\label{canonical_pm_C3}
\end{figure}

The first step is to identify the perfect matching that describes the crystal before any melting. For $\mathbb{C}^3$, this configuration is often referred to as the {\it empty room configuration} for its similarity with the empty corner of a $3d$ room. The corresponding perfect matching, which we will call the {\it canonical perfect matching} $p_0$,  is shown in \fref{canonical_pm_C3} \cite{Ooguri:2009ijd}. The figure shows a finite region that should be extended to the infinite hexagonal lattice in the obvious way. To simplify the visualization, we have omitted the black and white nodes of the brane tiling. The canonical perfect matching consists of three regions. Inside each of them, $p_0$ is given by one of the perfect matchings of the original brane tiling associated to one of the corners of the $\mathbb{C}^3$ toric diagram ($p_x$, $p_y$ and $p_z$). Interestingly, the boundary between these regions, represented with dotted lines in \fref{canonical_pm_C3}, agrees with the $(p,q)$ web in \fref{toric_C3_with_pms} (up to an obvious transformation). Brane tilings give rise to a ``discretized" version of the underlying CY$_3$ geometry \cite{Okounkov:2003sp,Ooguri:2009ijd,Chuang:2009crq,Eager:2011ns}. This fact holds for general toric CY$_3$'s. In \fref{canonical_pm_C3}, we have indicated the hexagons on which two or three regions coincide in blue. On these hexagons, $p_0$ is given by the appropriate combination of the basic perfect matchings.

\paragraph{Depth.}

\begin{figure}[H]
	\centering
	\includegraphics[height=5cm]{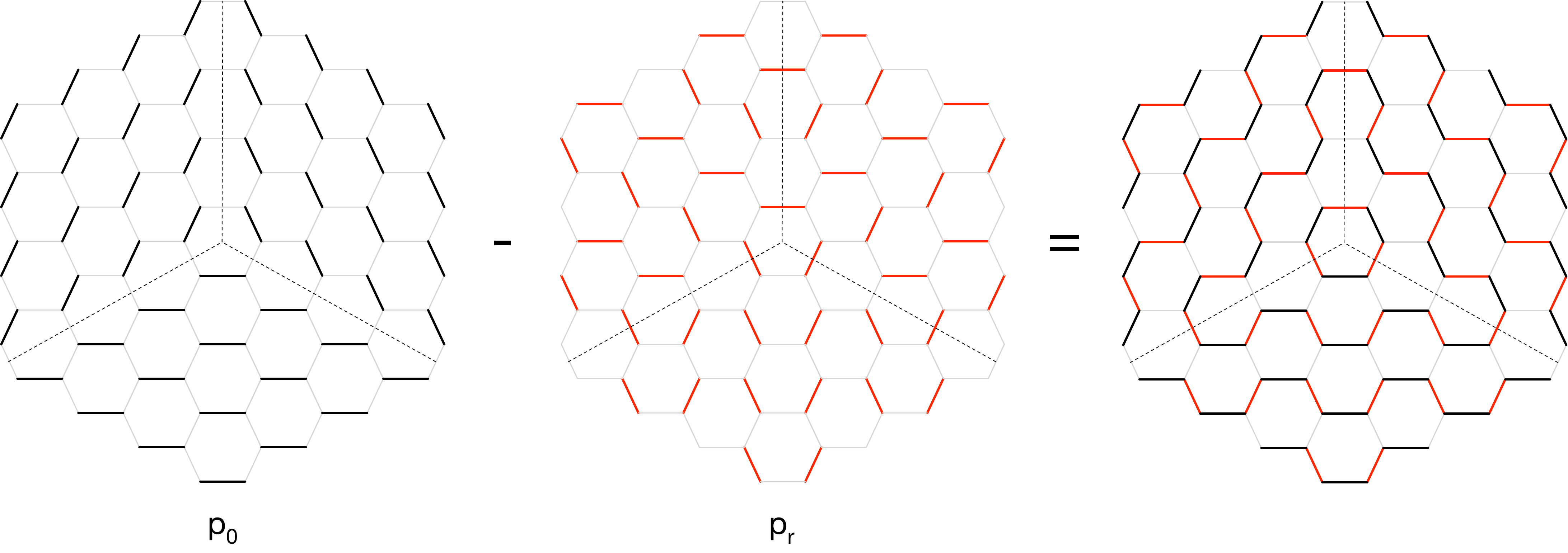}
\caption{The canonical perfect matching $p_0$, the reference perfect matching $p_r$ and the contour lines obtained from taking their difference.}
	\label{canonical-reference_pm_C3}
\end{figure}

The third dimension of the crystal is obtained by introducing a reference perfect matching $p_r$ and determining the height function. The procedure is analogous to the one for brane brick models discussed in Section \sref{section_height_function} (see e.g. \cite{Franco:2005rj} for details). \fref{canonical-reference_pm_C3} shows he choice of $p_r$ that, upon subtraction from $p_0$, reproduces the depth function measured from the top atom. Roughly speaking, within each of the three regions associated to one of the basic perfect matchings, $p_r$ is given by a combination of the other two. We also show the contour lines obtained from $p_0-p_r$. It is trivial to determine the orientation of these curves by taking into account the nodes of the brane tiling and their colors. 

\begin{figure}[H]
	\centering
	\includegraphics[height=5cm]{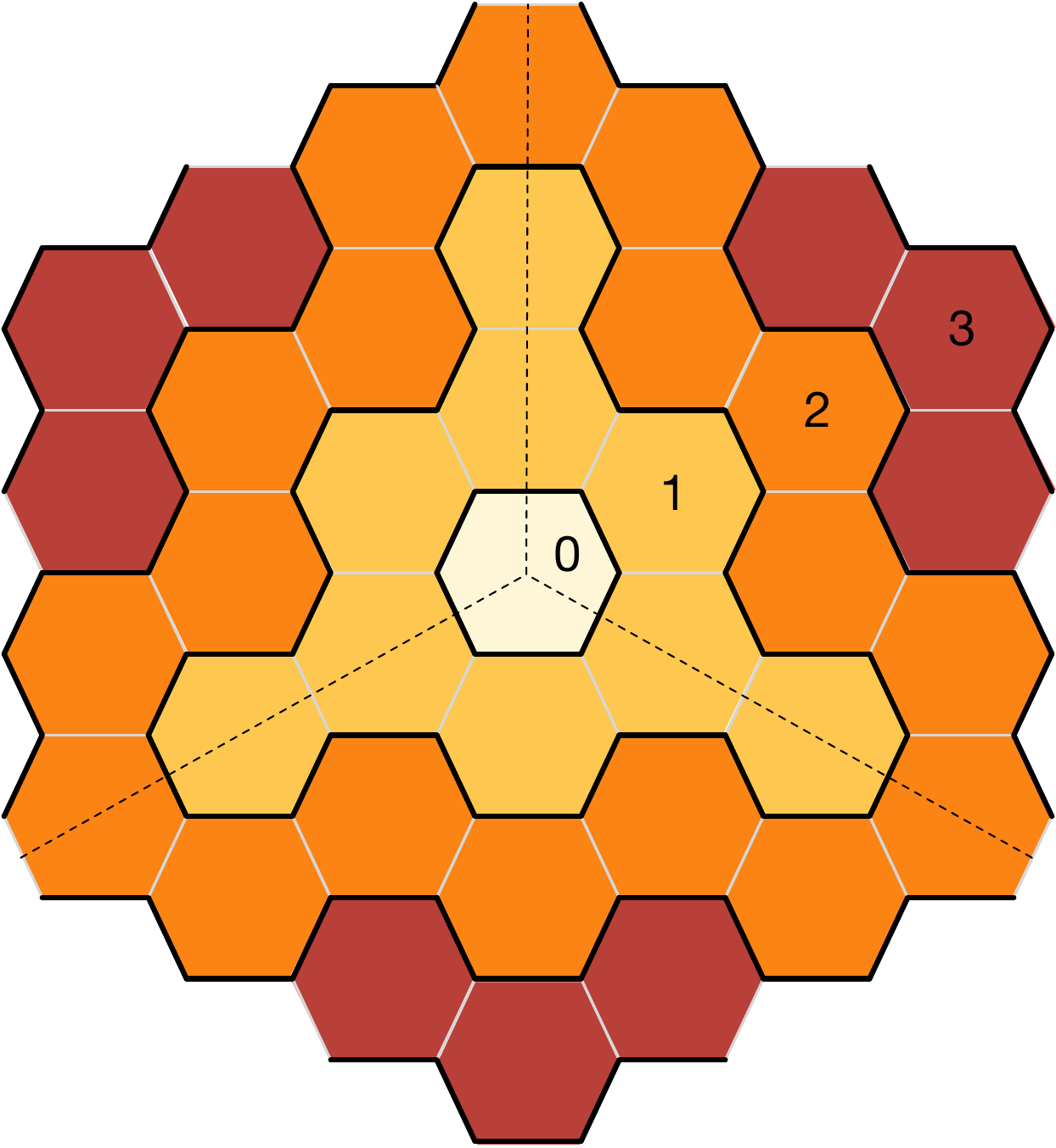}
\caption{Depth function for the canonical perfect matching, obtained from \fref{canonical-reference_pm_C3}.}
	\label{canonical-reference_pm_C3_height_function}
\end{figure}

It is natural to assign $d=0$ to the top atom and pick the orientation of the contours such that $d$ increases from there. The height function can then be regarded as the depth. We can interpret \fref{canonical-reference_pm_C3_height_function} as the corner of a cube, in which each of the three orthogonal faces correspond to one of the perfect matchings at the corners of the toric diagram of  $\mathbb{C}^3$ and the dashed lines indicate the edges at which two faces come together.  Notice that if we think that each of these faces is perpendicular to the $x$, $y$ and $z$ directions, our choice of $p_r$ is such that the depth increases in the $(1,1,1)$ direction. It is straightforward to verify that the resulting crystal agrees with the one built from the quiver, using the CY$_3$ analogue of the construction in Section \sref{section_model_crystal_melting_CY4_quiver} \cite{Ooguri:2009ijd}.

\subsection{Melting configurations}

\label{section_melting_configurations_CY3}

Let us now consider general melting configurations which, in this language, correspond to perfect matchings of the universal cover of the brane tiling. 

\paragraph{Depth.}

\begin{figure}[ht!]
	\centering
	\includegraphics[height=5cm]{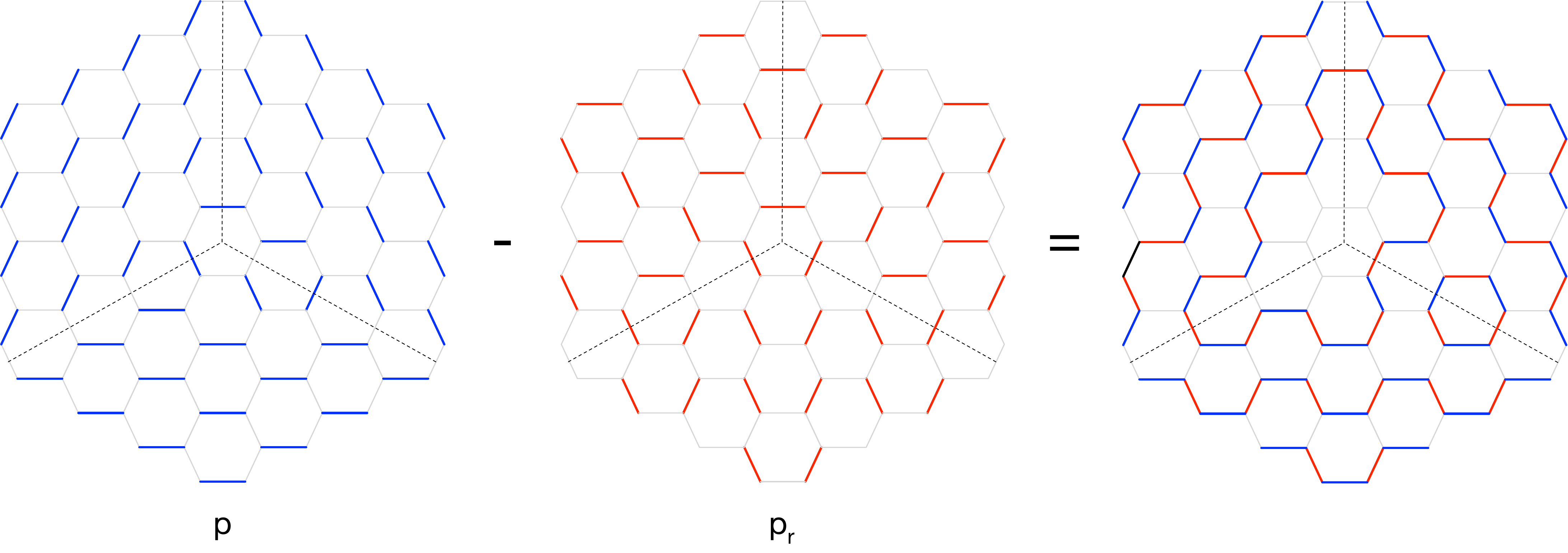}
\caption{A perfect matching $p$ and the contour lines of the height function obtained by subtracting the reference perfect matching $p_r$.}
	\label{pp-reference_pm_C3}
\end{figure}

\begin{figure}[ht!]
	\centering
	\includegraphics[height=5cm]{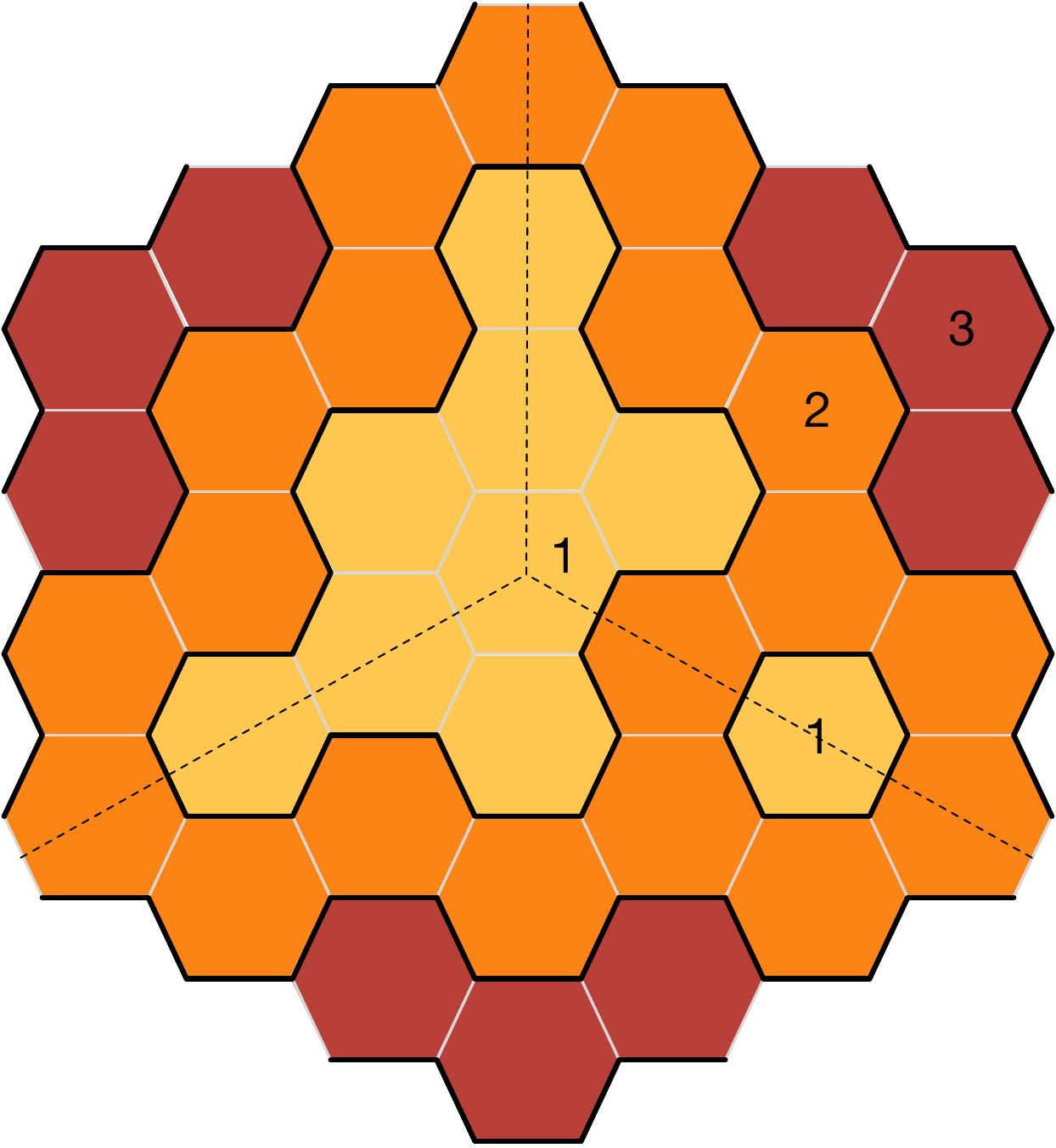}
\caption{Depth function for the perfect matching $p$ in \fref{pp-reference_pm_C3}.}
	\label{pp-reference_pm_C3_height_function}
\end{figure}

As for the canonical perfect matching discussed in the previous section, the depth function for any perfect matching $p$ is obtained by computing its difference $p-p_r$. Figures \ref{pp-reference_pm_C3} and \ref{pp-reference_pm_C3_height_function} show an example of this procedure for a general perfect matching $p$ and the resulting height function.

\paragraph{Melting height.}

As discussed in Section \sref{section_melting_configurations} for CY 4-folds, the atoms removed from the crystal are better captured by the melting height $h$, i.e. difference between the depths for $p$, given by \fref{pp-reference_pm_C3_height_function}, and for the unmolten crystal, given by \fref{canonical-reference_pm_C3_height_function}. In fact, it is not necessary to calculate the depth as an intermediate step, since the contour lines for $h$ can be computed directly as the difference $p-p_0$, i.e. they are independent of the reference perfect matching $p_r$. Finally, it is natural to assign $h=0$ to the region close to infinity, where $p$ and $p_0$ agree.

\begin{figure}[ht!]
	\centering
	\includegraphics[height=5cm]{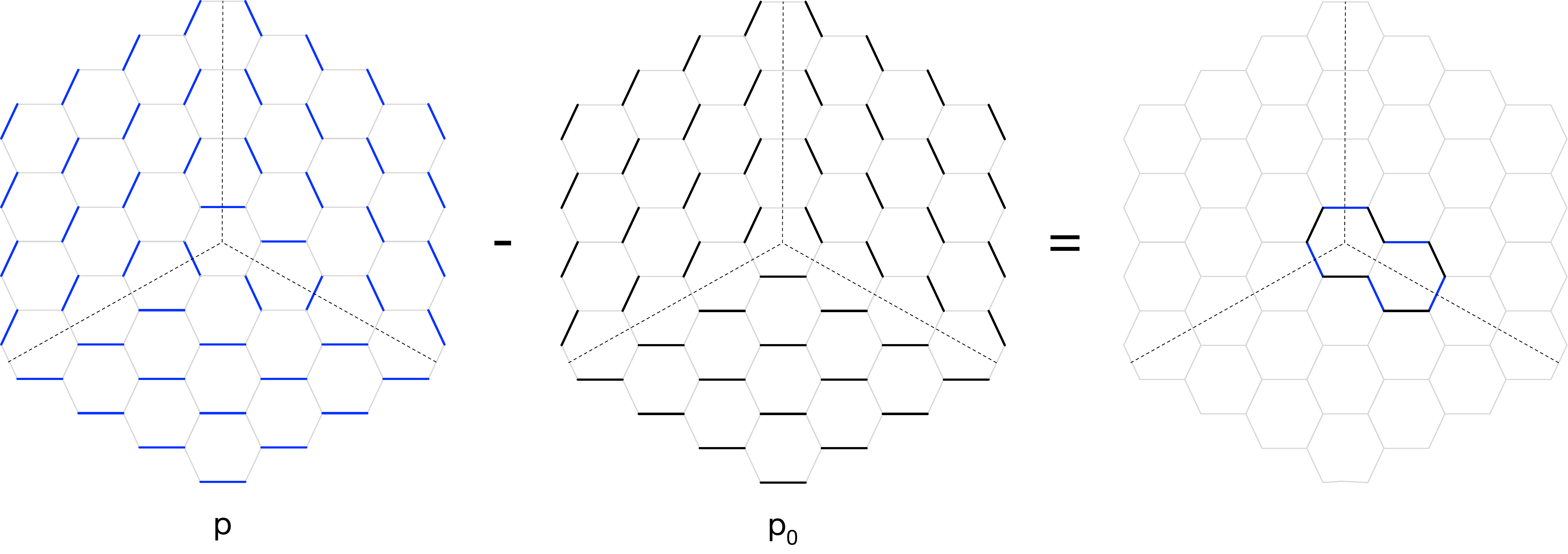}
\caption{A perfect matching $p$ and the contour lines of the melting height function obtained by subtracting the canonical perfect matching $p_0$.}
	\label{pp-p0_pm_C3}
\end{figure}

\begin{figure}[ht!]
	\centering
	\includegraphics[height=5cm]{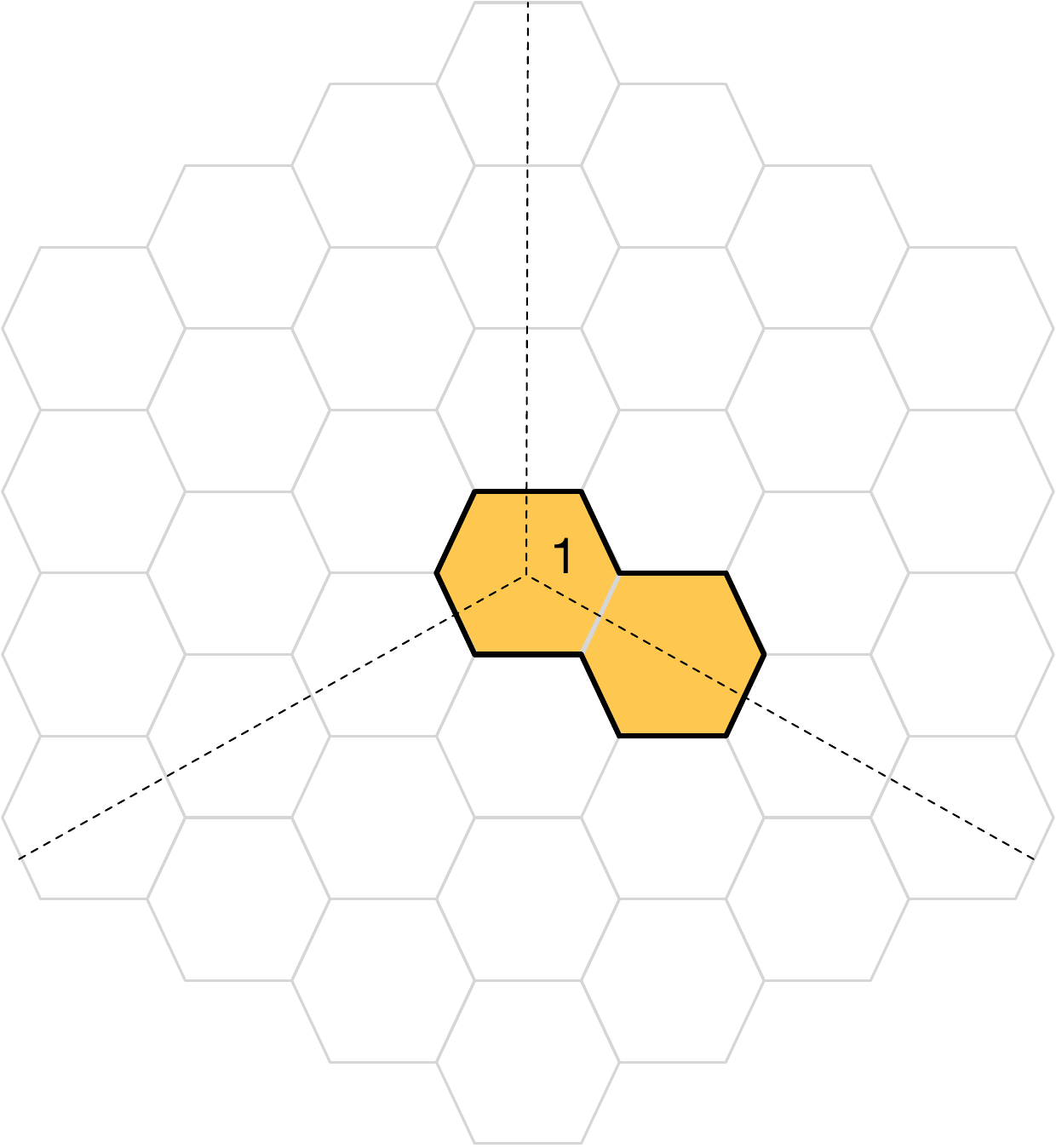}
\caption{Melting height function for the perfect matching $p$ in \fref{pp-p0_pm_C3}.}
	\label{pp-p0_pm_C3_height_function}
\end{figure}

Figures \ref{pp-p0_pm_C3} and \ref{pp-p0_pm_C3_height_function} show the determination of the melting height for the same perfect matching considered in \fref{pp-reference_pm_C3_height_function}. In this perspective, it is much cleared that the melting configuration under consideration corresponds to removing two atoms, the top one and another one along one of the three ridges of the crystal.

\section{Brane brick model description of CY$_4$ crystal melting}

\label{section_crystal_melting_BBM}

In this section, we explain how the melting model introduced in Section \sref{section_model_crystal_melting_CY4_quiver} in terms of periodic quivers can be formulated in the language of brane brick models and brick matchings. The classification of melting configurations translates into a problem of counting perfect matchings. While we illustrate our ideas with $\mathbb{C}^4$, they extend to general toric geometries \cite{WIP}. The discussion is a natural generalization of the one presented in Section \sref{section_appetizer} for CY 3-folds. 

\subsection{From brick matchings to melting configurations}

Melting configurations are in one-to-one correspondence with brick matchings of the universal cover of the brane brick model. We will elaborate on the details of this correspondence in Sections \sref{section_unmolten_crystal} and \sref{section_melting_configurations_BBM}. We now discuss how the fourth dimension of the crystal, i.e. the depth or the melting height, emerges from brick matchings.

\paragraph{Depth.}

Generalizing what happens for CY 3-folds, the fourth dimension of the crystal is obtained by introducing a reference brick matching $p_r$. For every brick matching $p$, the difference $p-p_r$ gives rise to collection of oriented {\it level surfaces}. The depth associated to these level surfaces is a special case of the height function discussed in Section \sref{section_height_function}. Below we explain how to identify the reference perfect matching $p_r$ that gives rise to the crystal defined by the quiver.

\paragraph{Melting height.}

The atoms removed from the crystal are more directly captured by the melting height $h$, the difference between the depths of the unmolten crystal and a melting configuration for any point $(x,y,z)$ in the quiver/brane brick model space. Generalizing the discussion in Section \sref{section_melting_configurations_CY3}, the level surfaces for $h$ in the melting configuration associated to a brick matching $p$ can be computed directly as the difference $p-p_0$, where $p_0$ is the {\it canonical brick matching}, without using the reference brick matching $p_r$. Finally, we set $h=0$ in the asymptotic region at infinity, where brick matchings agree with $p_r$.

\subsection{The unmolten crystal}

\label{section_unmolten_crystal}

\begin{figure}[ht!]
	\centering
	\includegraphics[height=10cm]{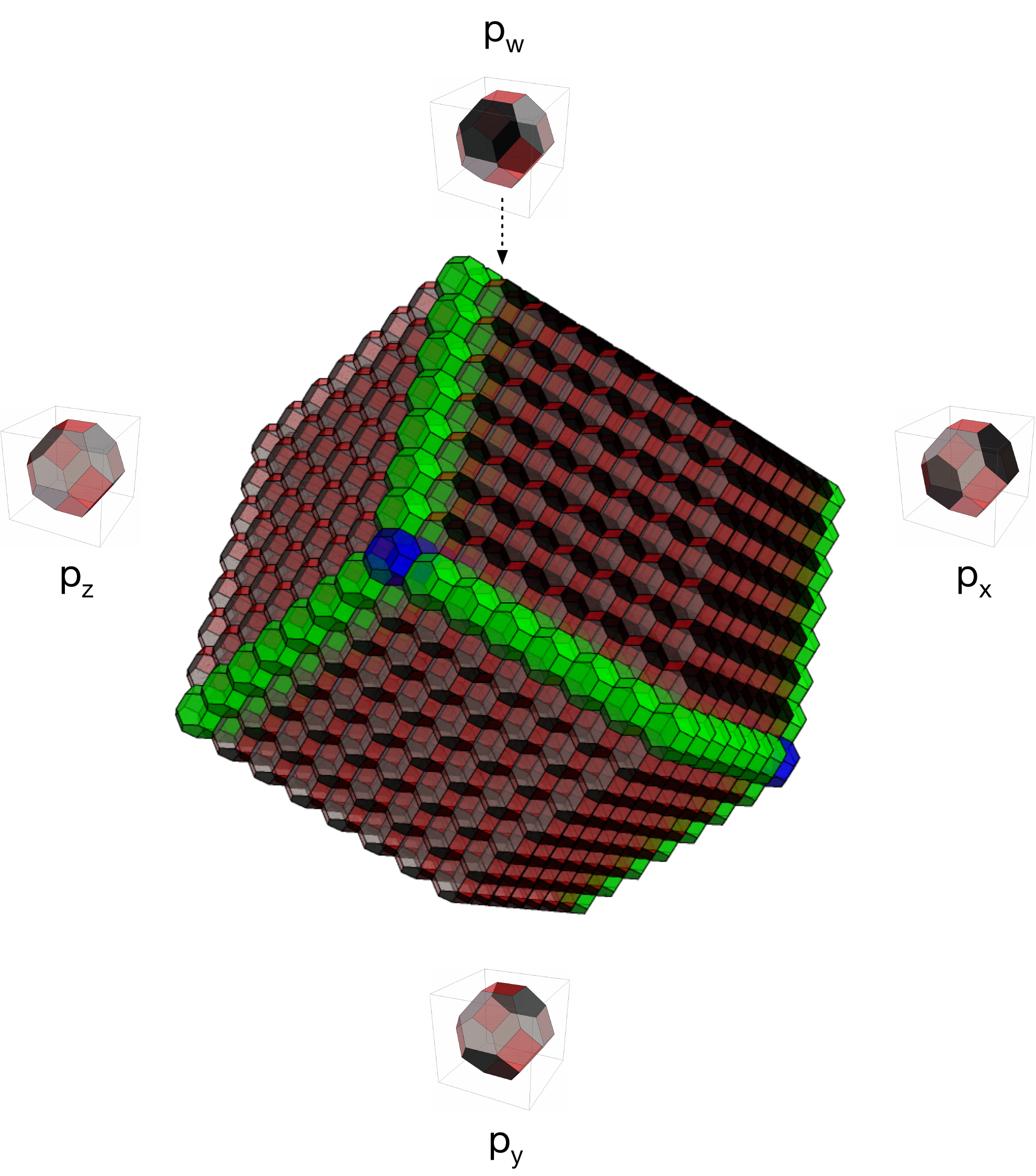}
\caption{Canonical brick matching for $\mathbb{C}^4$. It consists of four regions in which it coincides with the brick matchings associated to corners of the toric diagram. Bricks on which two or three of these regions overlap are shown in green and blue, respectively.}
	\label{canonical_pm_C4}
\end{figure}

\fref{canonical_pm_C4} shows the canonical brick matching $p_0$ of the universal cover of the brane brick model for $\mathbb{C}^4$ that describes the unmolten crystal. Generalizing the discussion in Section \sref{section_unmolten_crystal_C3}, $p_0$ consists of four regions, inside each of which it is given by one of the brick matchings of the brane brick model associated to a corner of the $\mathbb{C}^4$ toric diagram. The boundaries between every pair of these regions are six 2-dimensional surfaces that are in one-to-one them correspondence with the six edges connecting corners in the toric diagram. In other words, we can regard these surfaces as a ``$(p,q,r)$-web" dual to the toric diagram of a CY$_4$. We have indicated the bricks on which two or three regions coincide in green and blue, respectively. There is also a brick at the center of the configuration, and hence not visible in the figure, on which the four regions come together. On all these special loci, brick matchings are given by the appropriate combinations of the basic brick matchings.\footnote{The region shown in \fref{canonical_pm_C4} should be regarded as a finite subset of the {\it infinite} universal cover of the brane brick model. Its beautiful rhombic dodecahedron shape is an artifact of how it was generated, which is analogous to how the hexagonal region for $\mathbb{C}^3$ in the figures of Section \sref{section_appetizer} was created. In the normalization used in this figure, the universal cover of the brane brick model consists of bricks on a lattice generated by the following vectors: $v_x = (0, -2, \sqrt{2})$, $v_y = (0, 2, \sqrt{2})$, $v_z = (2,0, -\sqrt{2})$ and $v_W=(-2, 0,-\sqrt{2})$. Each of the regions covered by one of the basic brick matchings $p_i$, with $i=x,y,z,d$, are given by bricks located at  $n_j v_j + n_k v_k + n_l v_l$, with $j,k,l \neq i$ and $n_j,n_k,n_l=1 \ldots,n_{max}$, for some maximum size $n_{max}$. Also, the reason why the facets of the region in \fref{canonical_pm_C4} are not triangles as the ones in \fref{C4_stones_vs_depth} is simply that they do not correspond to surfaces of equal depth. \label{footnote_generators_BBM}} This perfect matching represents the $4d$ empty room configuration.

\begin{figure}[ht!]
	\centering
	\includegraphics[height=8cm]{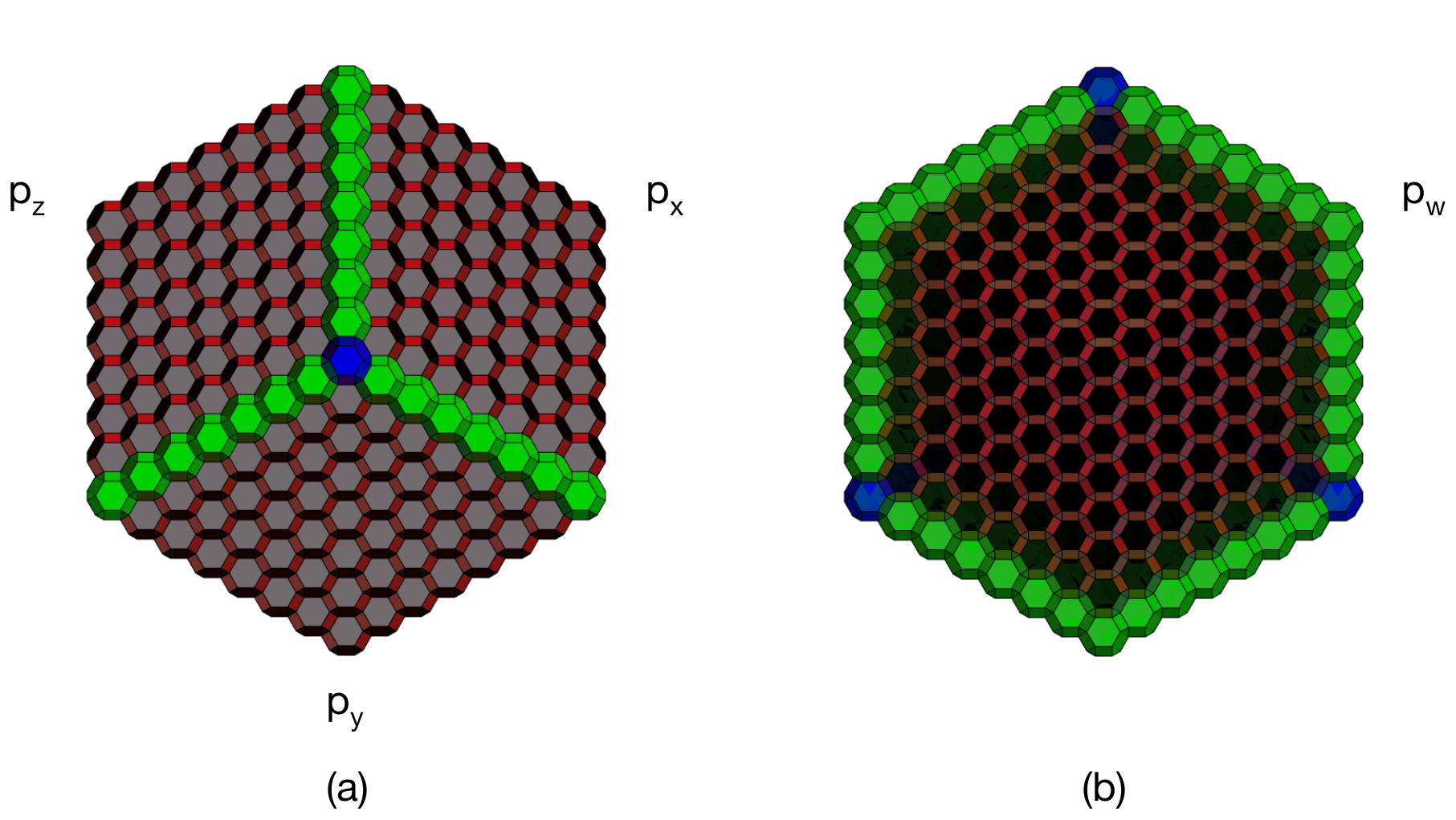}
\caption{a) When viewed from one of the $v_i$ vectors (in this case $v_w)$, the canonical brick matching for $\mathbb{C}^4$ reduces to the canonical perfect matching for $\mathbb{C}^3$. From the antipode, i.e. from the $-v_i$ direction, the configuration consists only of $p_i$ brick matchings.}
	\label{2_views_empty_corner}
\end{figure}

Let us explore this configuration in further detail. \fref{2_views_empty_corner}.a shows a view of $p_0$ along one of the blue rays, e.g. the vector $v_W$ in our construction. Remarkably, from this viewpoint, the configuration reduces to the canonical perfect matching for $\mathbb{C}^3$, which we presented in \fref{canonical-reference_pm_C3}. The truncated octahedra bricks and their brick matchings get projected onto the hexagonal lattice and its perfect matchings! \fref{2_views_empty_corner}.b shows a view from the antipode, i,e. from $-v_W$, from where we observe the entire plane covered by the remaining perfect matching $p_D$.

As for $\mathbb{C}^3$, the reference brick matching $p_r$ for $\mathbb{C}^4$ is basically given by the ``complement" of $p_0$ in each region, surface, etc. Since we will be primarily interested in the melting height of configurations, instead of the depth, and visualizing these perfect matchings is challenging, we will not present a figure with $p_r$.

\subsection{From crystal surfaces to brick matchings}

\label{section_from_surfaces_to_brick_matchings}

The canonical perfect matching we introduced above was determined such that, in combination with the reference brick matching, it gives rise to the depth determined by the quiver construction and illustrated in \fref{C4_stones_vs_depth}. This procedure translates into an algorithmic prescription for determining $p_0$ from the quiver, which we now review. The analogous construction for CY 3-folds was discussed in \cite{Ooguri:2009ijd}.

\begin{figure}[H]
	\centering
	\includegraphics[height=7.5cm]{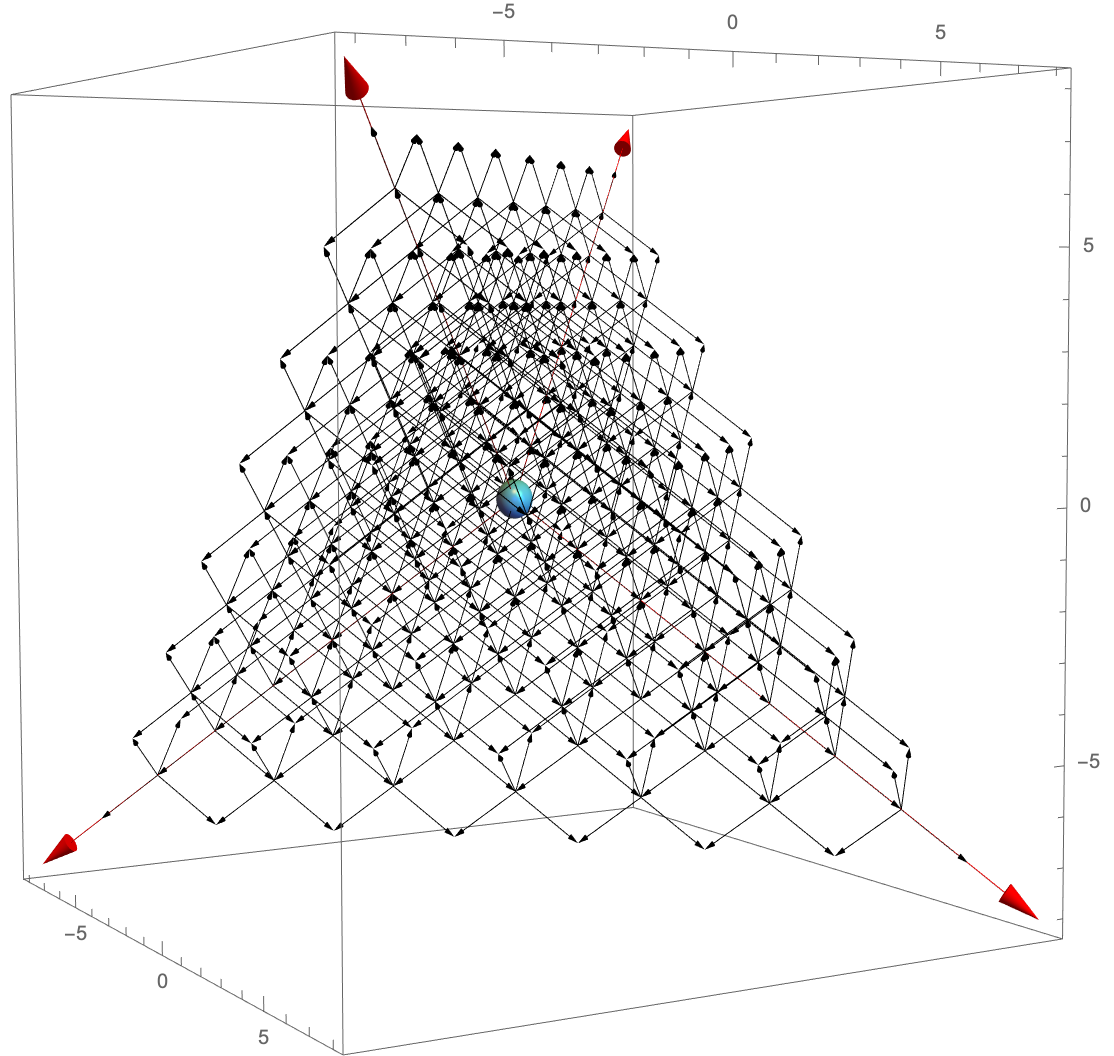}
\caption{Chiral arrows restricted to the atoms on the surface of the unmolten crystal up to $d=7$. The blue point indicates the atom at the origin.}
	\label{quiver_canonical_pm_C4}
\end{figure}

Let us focus on projection of the surface of the unmolten crystal onto quiver space and consider the chiral arrows connecting these atoms. Since we restrict to the surface of the crystal, these arrows do not form closed loops. If closed loops were present, they would give rise to atoms with the same coordinates on quiver space but different depths, therefore not on the surface. Equivalently, the arrows in this construction are those in the Hasse diagram of \fref{poset_C4} when restricted to atoms on the surface. The resulting quiver is shown in \fref{quiver_canonical_pm_C4}, where red arrows indicate the four primary directions of this crystal, $v_i$, $i=X,Y,Z,W$.  

From \fref{C4_stones_surface}, we know that the slices of the surface at each depth have a tetrahedral shape. When projected onto quiver space, the entire surface becomes an onion-like collection of tetrahedral shells of increasing size and depth. Since the arrows in \fref{quiver_canonical_pm_C4} point in the direction of increasing depth, they provide a quiver representation for the gradient of the depth. The canonical brick matching, once combined with the reference brick matching, gives rise to surfaces of increasing height (i.e. of decreasing depth). Therefore, we conclude that the arrows in the quiver in \fref{quiver_canonical_pm_C4} corresponds to $p_r$, while its complement correspond to $p_0$. The fact that the resulting objects are brick matchings follow from the absence of closed loops on this quiver and the definition of brick matchings in terms of chiral cycles given in Section \sref{section_brick_matchings}.

\bigskip

\paragraph{From melting configurations to brick matchings.}
The previous discussion can be extended to the surface $S_\mu$ of an arbitrary molten crystal $\mathcal{I}_\mu$ and result, following the same arguments, on a brick matching $p_\mu$. It therefore establish a correspondence between molten crystals or, equivalently, melting configurations and brick matchings.

\bigskip

To conclude this section, let us further scrutinize the connections between \fref{quiver_canonical_pm_C4} and our previous constructions. First of all, we observe that the vertices of the tetrahedral slices of the surface are along the $v_X$, $v_Y$, $v_Z$ and $v_W$ directions. This is easy to understand, since for any fixed depth $d$, the vertices correspond to the points of maximal distance from the origin. In turn, the distance is maximized by combining $d$ fields of the same type, hence aligning with one of the four primary directions.

Now consider a tetrahedral slice of the surface and focus on one of its triangular faces. The gradient of the depth function, which is represented by the arrows in \fref{quiver_canonical_pm_C4}, is orthogonal to these faces.\footnote{The notion of orthogonality is not perfectly defined in terms of the quiver, where we simply have collections of nodes, but it is clearer in terms of brane brick models.} For a given face, there are two normal vectors:
\begin{itemize}
\item $v_i$, which increases the depth along vertex $i$.
\item Since $v_X+v_Y+v_Z+v_W=0$, the other alternative is $-v_i=v_j+v_k+v_l$, where $j,k,l$ are the three remaining vectors. This vector points in the direction of increasing depth from each of the facets. This fact agrees with \fref{quiver_canonical_pm_C4}, where we observe that every face is associated to three types of arrows.
\end{itemize}
Recalling that $p_0$ is the complement of this quiver, these two observations are in perfect agreement with \fref{2_views_empty_corner}, where 
\begin{itemize}
\item From the $v_i$ direction, $p_0$ consists of three regions associated to $p_j$, $p_k$ and $p_l$ coming together.
\item From the $-v_i$ direction, $p_0$ corresponds entirely to $p_i$.
\end{itemize}

\subsection{Melting configurations and brick matchings}

\label{section_melting_configurations_BBM}

In this section, we discuss how general melting configurations are described in terms of brane brick models and brick matchings. For concreteness, we illustrate our ideas using Example 3 of Section \sref{section_pyramid_partitions}.

\begin{figure}[ht!]
	\centering
	\includegraphics[width=\textwidth]{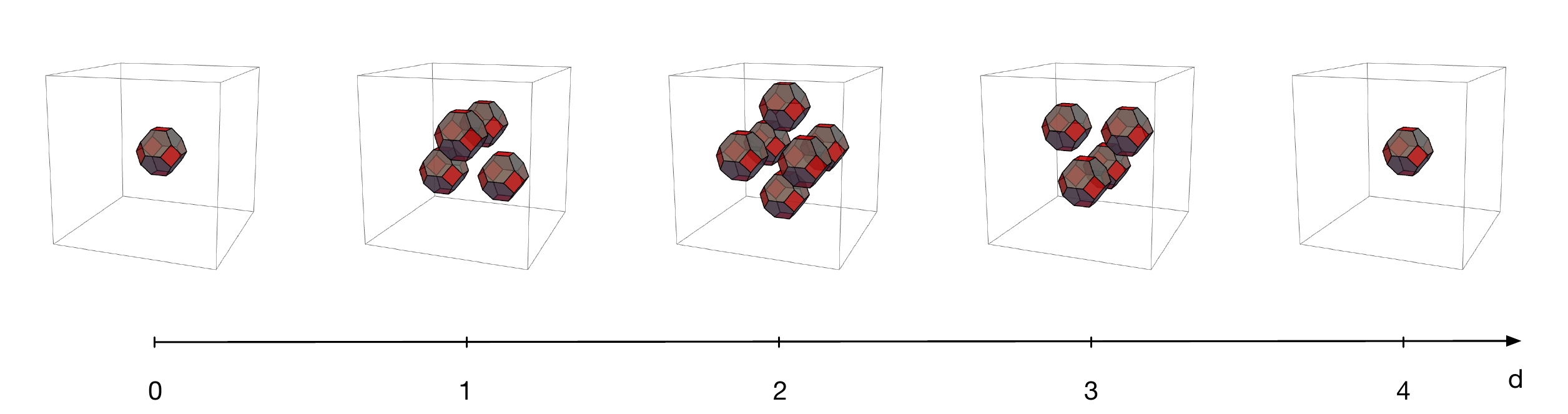}
\caption{Constant depth slices of the melting configuration defined by \fref{poset_height_2_configuration} in terms of the brane brick model.}
	\label{C4_16_stones_sliced_BBM}
\end{figure}

\begin{figure}[ht!]
	\centering
	\includegraphics[height=4cm]{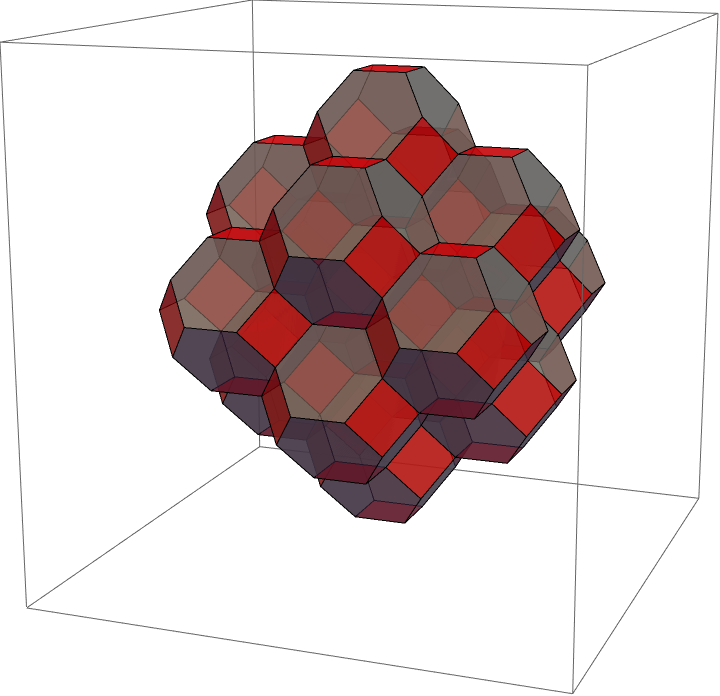}
\caption{Bricks in the universal cover of the brane brick model involved in the melting configuration under consideration. This region is the projection of the slices in \ref{C4_16_stones_sliced_BBM} onto quiver space.}
	\label{C4_16_stones_together_BBM}
\end{figure}

To start, let us visualize how the atoms in this partition translate into brane brick models. \fref{C4_16_stones_sliced_BBM} shows the constant depth slices of this melting configuration, originally shown in \fref{C4_16_stones_sliced}, in terms of the brane brick model.\footnote{Figures \ref{C4_16_stones_sliced_BBM} and \ref{C4_16_stones_sliced} differ by a $45^\circ$ rotation around the vertical axis. This follows from the different generators for the lattice chosen in each case. They are given in \eref{generating_vectors_quiver} and footnote \ref{footnote_generators_BBM}. It is a straightforward exercise to connect the two constructions.} \fref{C4_16_stones_together_BBM} shows the projection of these atoms onto quiver space. 

\begin{figure}[ht!]
	\centering
	\includegraphics[height=5cm]{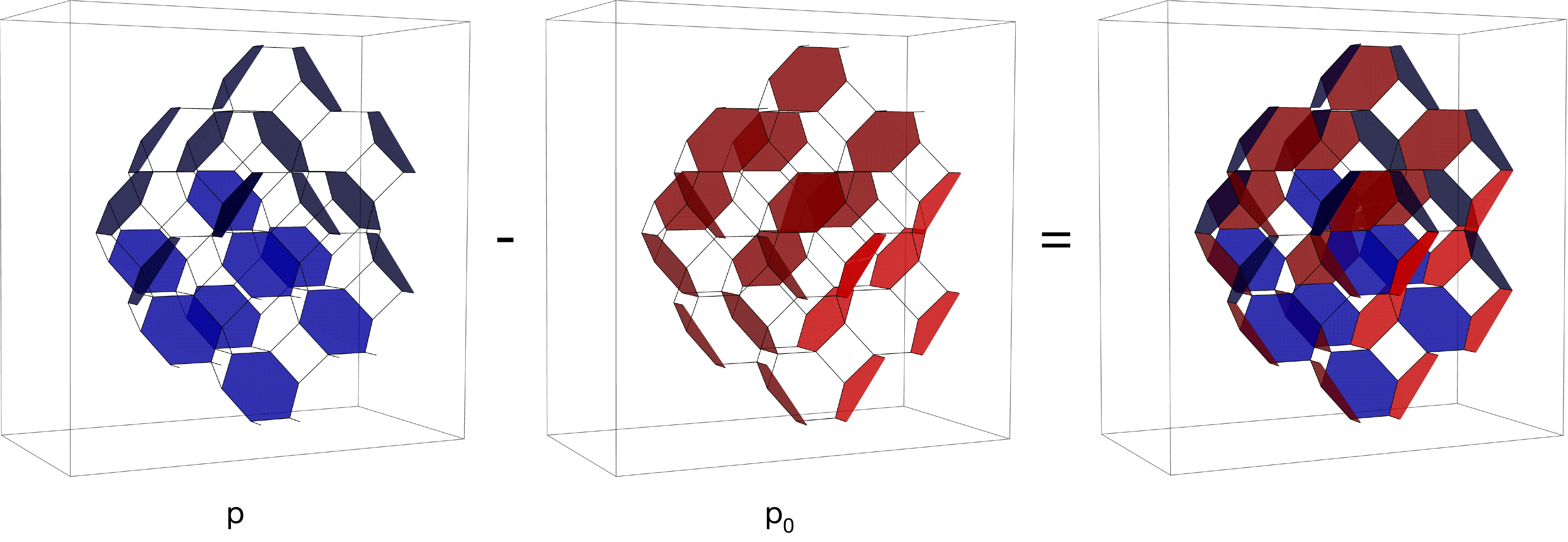}
\caption{A brick matching $p$, the canonical brick matching $p_0$ and the level surfaces of the melting height obtained by taking $p-p_0$.}
	\label{p-pr_C4}
\end{figure}

The brick matching $p$ corresponding to the melting configuration under consideration is shown in blue in \fref{p-pr_C4}, where we also show the canonical brick matching $p_0$ in red. We only give the chiral field content of both brick matchings, since Fermis can be reconstructed from this information. To simplify the figure, we restrict both brick matchings to the relevant region in \fref{C4_16_stones_together_BBM}. $p$ and $p_0$ coincide outside of this region and, therefore, they cancel out when subtracted. To be able to peek inside the configuration, we have split it open through the middle. The other halves of these brick matchings are identical the ones shown.

\begin{figure}[ht!]
	\centering
	\includegraphics[height=5cm]{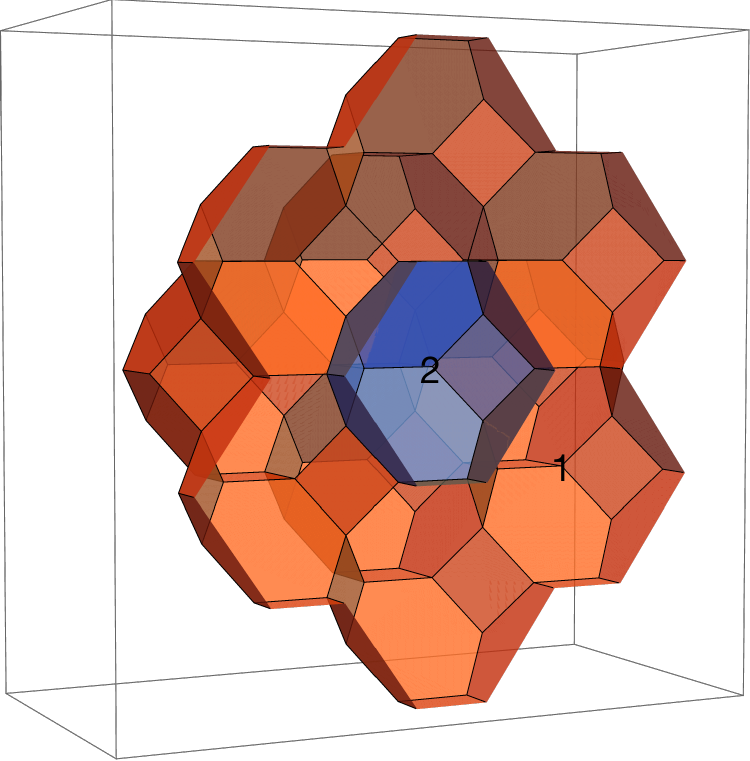}
\caption{Melting height for the brick matching $p$ in \fref{p-pr_C4}. The orange and blue regions have height 1 and 2, respectively.}
	\label{16_stones_BBM_height_2}
\end{figure}

The difference $p-p_0$ results into two nested oriented surfaces, as shown in \fref{p-pr_C4}. \fref{16_stones_BBM_height_2} shows the resulting melting height. The blue region at the origin has height 2, i.e. it corresponds to two overlapping atoms in the $4d$ crystal as expected. This is in perfect agreement with \fref{C4_16_stones_together}. The configuration has 14 bricks with melting height 1 and 1 brick with melting height 2, corresponding to 16 atoms, as expected.

\bigskip

It is worth noting that, from a graphing point of view, for general toric singularities it is often simpler to consider the periodic quiver version of crystal melting than its brane brick model counterpart, since representing the latter might be harder. Having said that, the brane brick model realization of the crystal melting model is conceptually important since, among other things, it maps the problem to the combinatorics of brick matchings.

\section{The general crystal melting model for toric CY$_4$'s}

\label{section_crystal_general_CY4}

While this paper focused on the D0-D8 system on $\mathbb{C}^4$, most of our discussion generalizes to arbitrary toric CY 4-folds and general brane configurations on them. Below, we briefly outline these generalizations and leave a detailed study to future work.

\paragraph{General CY 4-folds.}

For a general toric CY 4-fold, the underlying structure for the crystal is the universal cover of the corresponding periodic quiver $\tilde{Q}$.\footnote{Generically, there can be multiple periodic quivers, or equivalently brane brick modes, for a toric CY$_4$. They correspond to the so-called {\it toric phases} and are related by triality \cite{Gadde:2013lxa}. Such non-uniqueness is also present for CY 3-folds. We plan to study the crystals arising from different toric phases in future work.} The crystal has one type of atom for every node in the original quiver. We label each type of atom with an index $i$, with $i=1,\ldots, G$. $G$ is equal to the volume on the toric diagram normalized by the volume of a minimal tetrahedron.

\paragraph{General flavor branes.}

Let us first discuss the simplest configurations, i.e. those with single D8-brane wrapping the entire CY$_4$ with appropriate $B$-field, as considered in this paper. The D8-brane provides a single chiral flavor $q_{i_0}$ incoming into a node $i_0$ of the quiver. Generically, the resulting theory seems to depend on the choice of $i_0$, a freedom that is not present for $\mathbb{C}^4$. It would be interesting to investigate the dependence of the crystal melting models on $i_0$ and whether some criterion leads to a preferred choice. A similar freedom exists for CY 3-folds \cite{Ooguri:2009ijd}.

General flavor branes give rise to more involved configurations of flavors, consisting of $N_q$ incoming chirals $q_{i}$, $N_{\tilde{q}}$ outgoing chirals $\tilde{q}_{j}$ and $N_{\Psi}$ Fermis $\Psi_{k}$, where $i$, $j$ and $k$ indicate the nodes in the quiver to which the flavors are connected.\footnote{More broadly, one might consider crystals for more general flavor combinations, i.e. not necessarily associated to brane configuration. While less well motivated, such configurations might lead to interesting combinatorial problems.} $N_q,N_{\tilde{q}},N_{\Psi} \geq 0$ and, to keep the discussion general, we do not assume any relation between them. The flavors can participate in $J$- and $E$-terms, represented by gauge invariant terms of the following general forms:
\beq
q_{i} \mathcal{O}_{i,j} \Psi_{j}  \ \ , \ \ \overline{\Psi}_{i} \mathcal{O}_{i,j} \tilde{q}_j \ \ , \ \ q_{i} \Phi_{i,j} \tilde{q}_j \ \ , \ \  q_{i} \overline
{\Phi}_{i,j} \tilde{q}_j
\eeq
where $\mathcal{O}_{i,j}$ and $\Phi_{i,j}$ are operators made of D0-D0 fields. The $\mathcal{O}_{i,j}$ operators contain only chiral fields, $\Phi_{i,j}$ contain chirals and one Fermi, and $\overline{\Phi}_{i,j}$ contain chirals and a conjugate Fermi. These interactions should be added to the $J$- and $E$-terms that only involve D0-D0 fields, i.e. those encoded in the periodic quiver/brane brick model.

Motivated by the crystal models for CY 3-folds (see e.g. \cite{Ooguri:2009ijd,Chuang:2009crq,Eager:2011ns}), we propose that every atom in the unmolten crystal corresponds to an open oriented path of chirals starting at a $q_i$ modulo $J$- and $E$-term relations of both the D0-D0 Fermi fields and the Fermi flavors $\Psi_j$. Notice that some atoms might be reached by equivalent paths starting at different $q_i$'s. The positions of atoms in the crystal are determined by the rules in Section \sref{section_statistical_model_unmolten}. The crystal constructed in this way contains $N_q$ top atoms and is subject to up to $2 N_{\Psi}$ additional relations coming from the $\Psi_j$ fields. Finally, melting configurations are given by the melting rule of Section \sref{section_melting_configurations}.

While we are confident on the general picture outlined in this section, some of its details, particularly the proposed treatment of the $J$- and $E$-terms of $\Psi_j$ fields, deserve further study. 

Heuristically, we expect that if the number of relations coming from $J$- and $E$-terms for $\Psi_j$ fields exceeds the number of $q_i$ fields, this might lead to a truncation of the chiral operators associated to atoms. Therefore, depending on the relation between $N_q$ and $N_{\Psi}$ (and, possibly $N_{\tilde{q}}$), the resulting crystals might be infinite (like the one studied in this paper), finite, or infinite but effectively lower dimensional (like the example in Section \sref{section_D0-D6}). 
For similar phenomena for CY 3-folds, see \cite{Chuang:2009crq,Eager:2011ns}.

Finally, general flavor configurations can correspond to ``resolutions" of the discretized version of the toric CY$_4$ provided by the crystal, in which certain cycles grow to finite size. CY$_3$ examples displaying analogous behavior can be found in \cite{Chuang:2009crq,Eager:2011ns}. 

\medskip

\paragraph{Crystals and brane brick models.}

The implementation of the crystal melting model in terms of brane brick models and their brick matchings follows the general discussion of Section \sref{section_crystal_melting_BBM}. The starting point is the universal cover of the brane brick model for the toric CY 4-fold under consideration.

The brick matching $p_\mu$ associated to a melting configuration $\Omega_\mu$ is determined from the crystal surface $S_\mu$ using the method in Section \sref{section_from_surfaces_to_brick_matchings}. This includes the canonical perfect matching representing the unmolten crystal. Crystals associated to different flavor configurations, e.g. those corresponding to resolutions of the CY$_4$, are captured by different canonical perfect matchings.

For infinite crystals, every brick matching asymptotically approaches the unmolten crystal and contains a collection of ``frozen" regions inside each of which it is given by one of the brick matchings of the brane brick model associated to a corner of the toric diagram.

\medskip

\paragraph{Partition function.}

The partition function has a variable $y_i$, $i=1,\ldots,G$, for every type of atom in the crystal, i.e. for every gauge node in the quiver. It takes the form
\beq
Z=\sum_{\Omega_\mu} \prod_i y_i^{n^{(\mu)}_i} \, ,
\eeq
where the sum runs over melting configurations $\Omega_\mu$ and $n^{(\mu)}_i$ is the number of atoms of type $i$ in $\Omega_\mu$. For general CY 4-folds, every node in the quiver corresponds to fractional brane, which in turn is a bound state of D-branes wrapping vanishing cycles. Knowing these fractional branes, it is possible to re-express the partition function in terms of D-brane charges.

\subsection{A simple example: D0-D6 system on $\mathbb{C}^4$}

\label{section_D0-D6}

\begin{figure}[h]
	\centering
	\includegraphics[height=3.5cm]{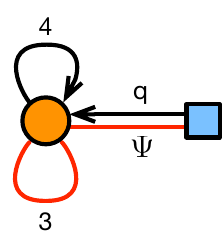}
\caption{Quiver diagram for D0-branes and a D6-brane on $\mathbb{C}^4$.}
	\label{quiver_C4_D0-D6}
\end{figure}

To illustrate the construction of crystals for more general flavor configurations, let us consider the case of D0-branes and a single D6-brane on $\mathbb{C}^4$. \fref{quiver_C4_D0-D6} shows the corresponding quiver. In the notation introduced above, this example has $N_q=N_\Psi=1$ and $N_{\tilde{q}}=0$.

Without loss of generality, let us assume that the D6-brane spans the $X$, $Y$ and $Z$ directions, and it is located at $W=0$. Then, in addition to the $J$- and $E$-terms of the D0-brane theory, which were given in \eref{EJ_C^4}, we have a $J$-term involving the flavors, given by the gauge invariant coupling $q \, W \Psi$. This coupling is easy to understand, since a (classical) expectation value for $W$ would result on a mass term for the flavor fields $q$ and $\Psi$, which agrees with the fact that separating the D6 and D0-branes along the $W$ direction makes the strings stretched between them massive.

\begin{figure}[ht!]
	\centering
	\includegraphics[width=\textwidth]{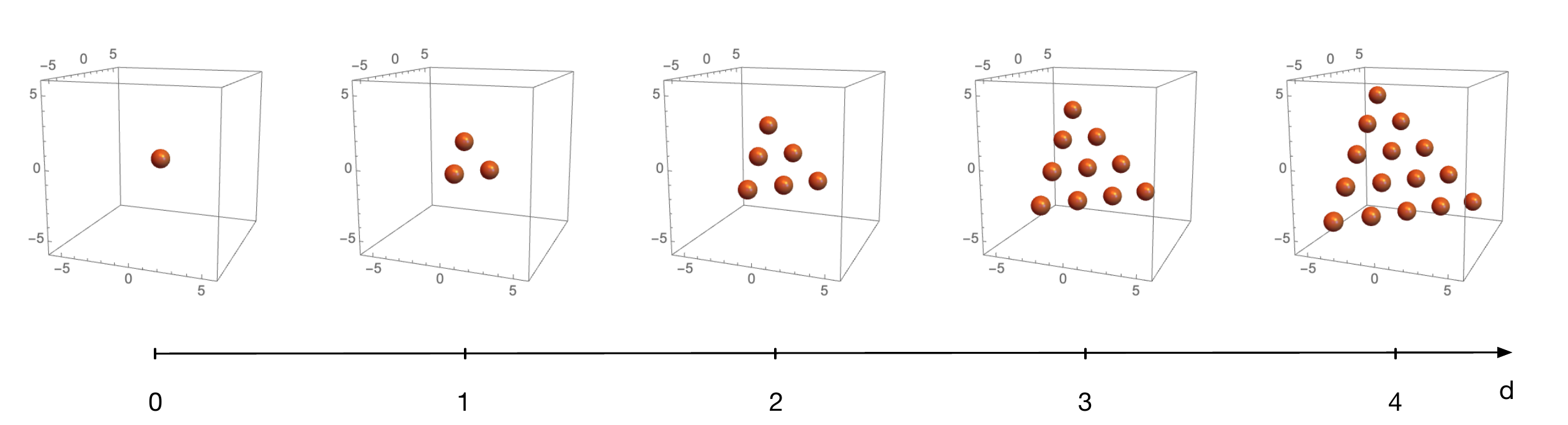}
\caption{Constant depth slices of the unmolten crystal for the D0-D6 system in $\mathbb{C}^4$ up to $d=4$.}
	\label{D0-D6_stones_vs_depth}
\end{figure}

The atoms in the unmolten crystal correspond to open oriented paths of chiral fields starting from $q$, modulo $J$- and $E$-term relations. If we considered only the $J$- and $E$-terms in \eref{EJ_C^4}, we would generate the same crystal we previously construceted in Section \sref{section_crystal_C4}. However, the vanishing of the $J$-term for $\Psi$ gives rise to the additional relation
\beq
q \, W =0 \, .
\label{J_term_Psi}
\eeq
Since every atom corresponds to a path containing $q$, \eref{J_term_Psi} implies that any path containing a $W$ field vanishes and the corresponding atom is not present in the crystal. Not surprisingly, we recover the well-known crystal melting model for D0-branes and a D6-brane on $\mathbb{C}^3$, which we also discussed in Section \sref{section_appetizer}. The crystal turns out to be infinite but 3-dimensional, with melting configurations in one-to-one correspondence with plane partitions. \fref{D0-D6_stones_vs_depth} shows constant depth slices of the unmolten crystal up to $d=4$. In contrast with \fref{C4_stones_vs_depth}, we observe that for this model the slices are 2-dimensional, in agreement with the fact that in this case the full crystal is 3-dimensional.

\section{Conclusions and future directions}

\label{section_conclusions}
 
We introduced a statistical model of crystal melting for non-compact toric CY 4-folds. While we focused on $\mathbb{C}^4$ to illustrate our construction, we discussed how it extends to general toric CY 4-folds. First, we implemented the model in terms of periodic quivers. We then reformulated it in terms of brane brick models and brick matchings. We introduced various techniques for visualizing the resulting crystals and their melting configurations, including slicing and Hasse diagrams. The crystals provide a discretized version of the underlying toric geometries. This paper takes the first steps laying out the basic ideas in what we consider will turn out to be a rich subject. Our work suggests various interesting directions for future investigation, some of which are summarized below:
 
 \begin{itemize}
 
\item We outlined the generalization of the melting crystal model to arbitrary toric CY 4-folds with general flavor D-branes. In the future, we plane to elaborate on this definition in further detail, explore such general setups and investigate whether they exhibit novel features. 
 
 \item Generically, there are multiple toric phases for a toric CY 4-fold. From a field theory perspective, such phases are related by {\it triality}, which is an IR equivalence of $2d$ $(0,2)$ gauge theories when the quiver theories are interpreted as living on D1-branes probing the CY$_4$ \cite{Franco:2016nwv,Franco:2016qxh}. In such cases, there are alternative structures underlying the crystal, given by the corresponding periodic quivers or brane brick models. CY 3-folds display a similar behavior, where toric phases are related by Seiberg duality \cite{Seiberg:1994pq,Feng:2000mi,Feng:2001xr,Beasley:2001zp,Feng:2001bn,Franco:2005rj}. We expect different phases to give rise to the same discretized toric geometry. It would be interesting to investigate the connection between crystals for different toric phases. 
 
\item Triality can change the framing flavors, even if the underlying periodic quiver remains the same. This transformation would result in crystals of different shapes and sizes. An analogous phenomenon has been studied for CY 3-folds (see e.g. \cite{Chuang:2009crq,Eager:2011ns}). In that case, Seiberg duality connects crystals, both infinite and finite, with different resolution parameters and sizes. These crystals encode the BPS spectrum within different stability chambers, which connect across walls of marginal stability. Interestingly, for CY 3-folds, the partition functions of crystals that are connected in this way transform as the variables of a cluster algebra with coefficients \cite{fomin2006cluster, Fock_2008,kontsevich2008stability,derksen2010quivers, Plamondon_2011,10.1215/00127094-2142753,Eager:2011ns}. It would be interesting to investigate how the partition functions of CY$_4$ crystals associated to flavor configurations connected by triality are related.

 \item It would be interesting to investigate the geometry of the molten crystal in the high temperature limit, i.e. for melting configurations with a large number of atoms, and whether it is connected to the mirror CY$_4$.
 
 \item {\it Quiver Yangians} are a new class of infinite-dimensional algebras that act on BPS states of non-compact toric CY 3-folds \cite{Li:2020rij,Galakhov:2020vyb}. These BPS states correspond to D-branes wrapping holomorphic cycles on such CY 3-folds and are captured by crystal melting models based on brane tilings. It is therefore possible to bootstrap quiver Yangians from molten crystal configurations. It would be interesting to determine whether similar algebras exist for BPS states on toric CY 4-folds and, if so, whether they are connected to the crystal melting models introduced in this paper.

\item The open string sector of the topological B-model model on CY $(m + 2)$-folds is described by $m$-graded quivers with superpotentials \cite{Franco:2017lpa,Closset:2017yte,Closset:2018axq}. This correspondence extends to general $m$ the connection between CY $(m + 2)$-folds and gauge theories on the worldvolume of D$(5-2m)$-branes for $m = 0, \ldots , 3$. $m$-dimers, a new type of combinatorial objects that fully encode the $m$-graded quivers and their superpotentials in the case in which the CY $(m+2)$-folds are toric was introduced in \cite{Franco:2019bmx}. For $m=1$ and 2 these objects correspond to brane tilings and brane brick models, respectively. Generalizing the well-known $m = 1$ and 2 cases, $m$-dimers significantly simplify the connection between geometry and $m$-graded quivers. It is natural to expect that a generalization of crystal melting models for toric CY $(m + 2)$-folds exists and that it is based on $m$-dimers. It would be interesting to pursue this line of investigation.

 \end{itemize} 
 
 We plan to address these questions in forthcoming work.
 
\acknowledgments

We would like to thank Yang-Hui He, Eduardo Garc\'ia-Valdecasas, Nikita Nekrasov, Nicolo Piazzalunga and, specially, Xingyang Yu for enjoyable and useful discussions. We are also grateful to Dongwook Ghim, Azeem Hasan, Sangmin Lee, Rak-Kyeong Seong and Cumrun Vafa for earlier collaborations on related topics. This work is supported by the U.S. National Science Foundation grants PHY-2112729 and DMS-1854179.

\bibliographystyle{JHEP}
\bibliography{mybib}

\end{document}